\documentclass[12pt,preprint]{aastex}

\begin{document}

\title{The Effect of Protoplanetary Disk Cooling Times on the 
Formation of Gas Giant Planets by Gravitational Instability}

\author{Alan P.~Boss}
\affil{Department of Terrestrial Magnetism, Carnegie Institution
for Science, 5241 Broad Branch Road, NW, Washington, DC 20015-1305}
\authoremail{aboss@carnegiescience.edu}

\begin{abstract}

Observational evidence exists for the formation of gas giant planets on
wide orbits around young stars by disk gravitational instability, but the 
roles of disk instability and core accretion for forming gas giants on 
shorter period orbits are less clear. The controversy extends to
population synthesis models of exoplanet demographics and to hydrodynamical
models of the fragmentation process. The latter refers largely to
the handling of radiative transfer in three dimensional (3D) hydrodynamical
models, which controls heating and cooling processes in gravitationally
unstable disks, and hence dense clump formation. A suite of models using 
the $\beta$ cooling approximation is presented here. The initial disks have 
masses of 0.091 $M_\odot$ and extend from 4 to 20 AU around a 1 $M_\odot$
protostar. The initial minimum Toomre $Q_i$ values range from 1.3 to 2.7, 
while $\beta$ ranges from 1 to 100. We show that the choice of $Q_i$ is
equal in importance to the $\beta$ value assumed: high $Q_i$ disks can be
stable for small $\beta$, when the initial disk temperature is taken 
as a lower bound, while low $Q_i$ disks can fragment for high $\beta$.
These results imply that the evolution of disks toward low $Q_i$ must
be taken into account in assessing disk fragmentation possibilities, at
least in the inner disk, i.e., inside about 20 AU. The models suggest that 
if low $Q_i$ disks can form, there should be an as yet largely undetected 
population of gas giants orbiting G dwarfs between about 6 AU and 16 AU.

\end{abstract}


\section{Introduction}

 Two end-member mechanisms have been proposed for gas giant planet formation,
core accretion (e.g., Mizuno 1980) and gas disk gravitational instability (e.g., 
Cameron 1978). The former depends on the reliable mechanism of the collisional 
accumulation of solid bodies, widely accepted as the formation mechanism
for the terrestrial planets (e.g., Wetherill 1990, 1996). If a growing solid
core should exceed a critical core mass (Mizuno 1980), its atmosphere will
become unstable and the core will rapidly gain mass by accretion of gas
and solids from the surrounding disk (e.g., Pollack et al. 1996), resulting
in a gas giant planet. Disk instability, on the other hand, relies on
the disk being massive and cold enough to become gravitationally unstable,
forming spiral arms that grow and collide to form dense, self-gravitating 
clumps of gas and dust that might collapse and contract to form giant gaseous 
protoplanets (e.g., Boss 1997). Hybrid mechanisms may also be considered for 
forming terrestrial and gas giant planets (e.g., Boss 1998; Nayakshin 2010).
Lissauer \& Stevenson (2007) presented a review of the core accretion mechanism,
while Durisen et al. (2007) reviewed the disk instability mechanism. More
recently, Helled et al. (2014) reviewed both mechanisms, listing the strengths
and weaknesses of each end-member, without reaching a definite conclusion 
about either mechanism. Rather, Helled et al. (2014) called for considerably
more research on specific theoretical problems for both mechanisms. One
problem highlighted by Helled et al. (2014) is the question about whether 
or not disk instability can lead successfully to giant protoplanet 
formation inside about 20 AU, where high optical depths in the disk mean 
that radiative cooling is crucial for clump formation and survival. That
challenge is the focus of this paper.

 We begin with a brief summary of the current status of observational
constraints on giant planet formation models, both from studies of  
protoplanetary disks and from exoplanet detection surveys, before turning
to a reconsideration of the disk radiative cooling problem that is the
primary motivation for this new suite of disk instability models.

\section{Observational Constraints on Theoretical Models}

 Observational constraints on planet formation models range from estimates
of the initial conditions for planet formation, namely studies of
protoplanetary disks, which provide the feedstock for planetary systems, 
to exoplanet demographics, the final result of the planet formation process.
Disk masses, temperatures, and lifetimes are key discriminators for
comparing the disk instability and core accretion mechanisms.

\subsection{Protoplanetary Disk Masses}

 The most essential constraint is having sufficient disk mass to form a 
planetary system, by either core accretion or disk instability.
The minimum mass solar nebula was estimated by 
Weidenschilling (1977) to be in the range of 0.01 to 0.07 $M_\odot$,
but most citations refer only to the {\it minimum} of this range
(0.01 $M_\odot$) as being the minimum mass solar nebula. This
estimate also assumes a 100\% efficient planet formation process,
and so in reality these estimates of 0.01 to 0.07 $M_\odot$ represent
{\it lower bounds} on the mass of the protoplanetary disk needed to 
produce our solar system (Weidenschilling 1977). From this point of
view, the solar nebula might well have had a mass of $\sim 0.1 M_\odot$
or more.

 Considerably more information on the range of protoplanetary disk 
masses is obtained by millimeter-wave astronomy.
Millimeter wave observations of 11 low- and intermediate-mass 
pre-main-sequence stars implied that their disks formed with masses
in the range from 0.05 to 0.4 $M_\odot$ (Isella et al. 2009).
Submillimeter wave observations of T Tauri stars in the $\sim$ 1 Myr old
Ophiuchus star-forming region imply disk masses in the range from
0.004 to 0.143 $M_\odot$ (Andrews et al. 2010). Similar observations
of young stars in the even younger ($\sim$ 0.5 Myr old) cluster NGC 2024
found disk masses ranging from $\sim$ 0.003 to 0.2 $M_\odot$ 
(Mann et al. 2015), with $\sim$ 10\% having disk masses greater
than 0.1 $M_\odot$, a considerably higher fraction than in the
older Taurus star-forming region. Millimeter 
wave observations of Class II disks in the Taurus star-forming region 
imply that the inferred dust disk masses scale roughly proportionally to
the masses of their host stars (Andrews et al. 2013). However, estimates of
total disk masses have long suffered from the uncertain conversion of the 
observed mass of the growing dust grains (e.g., Banzatti et al. 2011)
that produce the submillimeter continuum emission to 
the inferred gas disk mass, potentially leading to substantial
underestimates of the total disk masses (Andrews \& Williams 2007). High optical
depth midplanes, even at millimeter wavelengths, can lead to weak
dust continuum emission and hence underestimates of the true 
disk mass (Forgan et al. 2016).
Disk mass estimates based on observations of molecular species also
depend on the assumed grain size distribution, leading to underestimated
disk masses by as much as a factor of 10 (Miotello et al. 2014;
Dunham et al. 2014). However, observations of the isotopologues of CO gas  
may offer a direct estimate of the mass of gas in protoplanetary disks 
(Miotello et al. 2016).

 Disks with masses of $\sim$ 25\% that of their
host star will be gravitationally unstable and produce spiral arms
similar to those in two protoplanetary disks, MWC 758 and SAO 206462
(Dong et al. 2015). The young stellar object (YSO) Elias 2-27, with a 
mass of $\sim 0.5 - 0.6 M_\odot$, has a protoplanetary disk with
a mass of $\sim 0.04 - 0.14 M_\odot$ and trailing, symmetric spiral
arms that extend to the disk midplane (P\'erez et al. 2016).
Episodic FU Orionis outbursts in young solar-type 
stars are best explained by disks that are at least partially 
gravitationally unstable (e.g., Zhu et al. 2007, 2009, 2010; 
Vorobyov \& Basu 2010a,b; Liu et al. 2016). 
Taken together, all of these observations suggest that a significant
fraction of protoplanetary disks are still gravitationally unstable
during their earliest phases of evolution, when massive protostellar disks
transition into less-massive protoplanetary disks.

\subsection{Protoplanetary Disk Lifetimes}

 Disk lifetimes are another key constraint, especially for the
core accretion mechanism, where gas disk lifetimes of $\sim$ 1 Myr
or longer are typically required to form gas giant planets (e.g., Pollack 
et al. 1996; Hubickyj et al. 2005). The conventional wisdom, based on IR
excesses, is that typical disk lifetimes are in the range of 1 to 10 Myr, with
an average age of $\sim$ 3 Myr (e.g., Figure 11 in Hernandez et al. 2008).
These statistics refer to young stars of different evolution classes
in a variety of stellar groups. When restricting the sample to include
only class III sources, Cieza et al. (2007) found that roughly 
$\sim$ 50\% of the youngest of 230 weak-line T Tauri stars
showed no evidence for IR excesses, suggesting that half of the disks 
dissipate on time scales of $\sim$ 1 Myr or less. These observations
refer solely to IR excesses, wavelengths where the disks are optically thick 
and the dust disk mass is difficult to estimate.
However, protoplanetary disks are optically thin at submillimeter wavelengths,
making submillimeter dust continuum observations capable of producing more 
reliable dust disk masses, and hence gas disk masses. A submillimeter survey
of nearly 300 YSOs in the $\sigma$ Orionis
cluster detected only 9 disks, finding a mean disk mass of 
0.5 Jupiter masses ($M_J$) for these YSOs with ages 
of $\sim$ 3 Myr (Williams et al. 2013), and implying that gas giant
planet formation needs to finish well within 3 Myr. A similar survey
of the slightly younger (2-3 Myr old) cluster IC 348 found disk masses
in the range of 1.5 to 16 $M_J$ (Cieza et al. 2015) for 13 out of
the 370 cluster members, implying that disks with at least the 
mass ($\sim 10 M_J$) of the lower bound on the minimum mass solar nebula 
(Weidenschilling 1977) are very rare, occurring less than about 1\% of the 
time at the age of the IC 348 cluster. In contrast, near-IR excesses
occur for roughly half of the stars in IC 348 and $\sigma$ Orionis
(e.g., Hernandez et al. 2008), showing that 
the conventional wisdom about gas disk lifetimes has been severely
skewed by a reliance on IR excesses. The submillimeter surveys imply
a typical gas disk lifetime closer to 1 Myr than to 3 Myr, at least
for disks capable of forming gas giant planets.

\subsection{Protoplanetary Disk Temperatures}

 Disk midplane temperatures, along with disk surface densities, determine
the Toomre (1964) $Q$ stability value, and hence the propensity for a disk to be
gravitationally unstable. The solar nebula is the protoplanetary disk for 
which one would think we have the best information about its properties.
Cometary compositions are often used to place constraints on solar
nebula temperatures at their time of formation. Molecular abundance
ratios determined for the comet 67P imply that if its ice grains agglomerated 
from clathrates (Mousis et al. 2016), then this Jupiter-family comet must
have formed in a region of the solar nebula with temperatures in
the range of about 32 K to 70 K (Lectez et al. 2015). Jupiter-family
comets are believed to have been formed beyond Neptune, in the Kuiper Belt,
whereas Oort Cloud comets are thought to have formed inward of Neptune, 
where Jupiter and the other giant planets could scatter them outward to
much longer period orbits. Comet C/1999 S4 is a long period comet from
the Oort cloud, containing ammonia ices with ortho-to-para ratios (OPR)
that imply formation at a temperature of about 28 K (Kawakita et al. 2001). 
Measurements of the OPR for cometary water have often been used to infer 
even lower nebular temperatures, in the range of 10 K to 20 K (e.g., 
Hogerheijde et al. 2011). However, Hama et al. (2016) have disputed 
the commonly assumed relationship between water OPR and temperature, casting
some doubt on use of the water OPR as a cometary thermometer for the solar
nebula. 

 Other disk temperature estimates come from observations of protoplanetary 
disks around young solar-type stars. Observations of the DM Tau outer disk, 
on scales of 50 to 60 AU, imply midplane temperatures of 13 to 20 K
(Dartois et al. 2003), with surface temperatures of $\sim$ 30 K. Theoretical
disk models that reproduce the observed spectral energy distributions
for T Tauri-star disks typically predict disk midplane temperatures of
$\sim$ 200 K at 1 AU, $\sim$ 40 K at 10 AU, and $\sim$ 15 K at 100 AU
(Lachaume et al. 2003; D’Alessio et al. 2006). These midplane temperatures 
are low enough that provided the disk is massive enough, the disk is
likely to be gravitationally unstable.

\subsection{Exoplanets Embedded in Protoplanetary Disks}

 The spectacular images of the HL Tau protoplanetary disk (ALMA Partnership
et al. 2015) show several gap-like structures centered on the 
$\sim 1.3 M_\odot$ central protostar. HL Tau is less than 1-2 Myr
old and has a disk mass of $\sim 0.03 - 0.14 M_\odot$. Models
suggest that the three main gaps at 15, 35, and 70 AU could be caused
by embedded protoplanets with masses of 0.2, 0.27, and 0.55 $M_J$
(Dipierro et al. 2015). A direct imaging search for embedded planets
in the 70 AU gap placed only upper limits of $\sim 10 - 15 M_J$ on
the unseen objects (Testi et al. 2015). Testi et al. (2015) noted
that Boss (2011) showed that the HL Tau disk could form a $\sim 5 M_J$
planet at 70 AU by disk instability, a planet massive enough to
cause the gap, but not massive enough to have been directly imaged.
Akiyama et al. (2016) have shown that all of the observed gaps in the
HL Tau disk are consistent with gas giant formation by disk instability,
coupled with inward migration. Embedded planets can also generate
spiral arms in marginally gravitationally unstable disks 
(Pohl et al. 2015). Carrasco-Gonzalez et al. (2016) have 
used the VLA to study the HL Tau disk, finding that the inner disk has 
fragmented and formed dense clumps, suggesting gravitational instability 
as the cause (see also Mayer et al. 2016).
The HL Tau system thus may represent the foremost poster 
child for gas giant planet formation by disk gravitational instability.

 The HL Tau disk is threaded with a magnetic field that is coincident
with its major axis (Stephens et al. 2014), implying that the field 
does not control the dynamics of the disk. In a related vein, recent work
on the Allende meteorite has shown that its magnetization was derived
from its parent body, not the solar nebula (Fu et al. 2014), implying
that the dynamics of the midplane of the solar nebula were not dominated 
by magnetic forces. Hence non-magnetic disk instability models appear
to be relevant to both the HL Tau disk and the solar nebula, and presumably
to other protoplanetary disks as well.

\subsection{Exoplanet Demographics}

The ultimate constraints on theoretical models of planetary formation 
are derived from the results of the ongoing world-wide effort to
determine the population statistics of exoplanets with widely varying
masses, mean densities, atmospheric compositions, and orbital properties.
The spectacular discoveries of thousands of exoplanets by the Kepler 
Mission (Borucki et al. 2010, 2011a,b) have revolutionized the field,
and made possible direct comparisons with the predictions of exoplanet
population synthesis (EPopS) models based on the core accretion mechanism.
Perhaps most dramatically, the Kepler detections imply that the 
number of planets per star appears to be a roughly monotonically increasing
function of decreasing exoplanet radius (Fressin et al. 2013), at 
least down to radii of $\sim 2 R_\oplus$ (Earth radii), contrary to the 
predictions of EPopS models. Early EPopS models (e.g., Ida \& Lin 2005, 2008) 
were based on the population uncovered primarily by radial velocity (RV) and 
ground-based transit photometry surveys and were able to select model
parameters that fit reasonably well the exoplanets known at the time.
These models (e.g., Figure 3 in Ida \& Lin 2008) predicted a dearth of 
planets with masses in the super-Earth mass range and short period orbits
(i.e., inside $\sim$ 1 AU), the so-called {\it exoplanet desert}
(Ida \& Lin 2004), as well as a 
surplus of hot Jupiters and hot super-Earths orbiting at $\sim$ 0.03 AU.
On the contrary, Kepler found an {\it exoplanet oasis} where the EPopS 
predicted a desert: Batalha (2014) showed that Kepler had found the
highest frequency of exoplanets in the region where the desert had
been predicted, and no evidence for a pile-up of hot exoplanets with
orbital periods of a few days. The same problem besets the more recent 
EPopS models of Alibert et al. (2011) and Mordasini et al. (2012). 
EPopS models with a new prescription for orbital migration
have filled in part of the desert, but only at the cost of over-producing 
massive gas giants and shifting the desert upward to higher masses
(see Figure 8 in Dittkrist et al. 2014; see also Coleman \& Nelson 2014). 
Evidently EPopS models based solely on core accretion are not yet able 
to find a combination of parameters that enables them to reproduce 
the Kepler findings. EPopS models that include gas giant planet formation
by disk instability (e.g., Forgan \& Rice 2013) might offer a means to 
escape some of the inward orbital migration and surplus gas giant planet
problems that beset pure core accretion EPopS models by shortening gas 
disk lifetimes, and by stopping cores from migrating inward and growing
to become gas giants.

 The Kepler Mission's 4-yr survey of the field in Cygnus-Lyra prohibited
the detection of exoplanets with orbital periods greater than about 1 yr.
Other techniques shed light on the demographics at greater distances.
E.g., an RV survey of 202 solar-type stars that had been
followed for 17 yrs found a frequency of $\sim$ 6\% for giant planets 
orbiting from 3 to 7 AU (Wittenmyer et al. 2016). The combination
of Adaptive Optics (AO) imaging and a long-term RV survey found a
total occurrence rate of $52 \pm 5$\% for exoplanet masses in the range of
1-20 $M_J$ at distances of 5 to 20 AU (Bryan et al. 2016), with a
suggestion that most have orbital distance less than 10 AU. This
fraction is higher than that found earlier by Cumming et al. (2008),
who found that 8 yrs of RV data implied that up to 20\% of stars
have gas giants within 20 AU. Evidently longer survey periods find
increasing frequencies of long-period exoplanets, as expected. Hence
even the current estimates should be considered lower bounds.

 Microlensing surveys also sample a different portion of exoplanet discovery
space compared to transit photometry surveys like Kepler, namely orbital
distances comparable to the Einstein radius, which means orbital 
distances of a few AU and orbital periods of a few
years for solar-type stars. Cassan et al. (2012) found
their microlensing data to reveal that $\sim$ 17\% of stars host planets
with masses in the range of 0.3 to 10 $M_J$, while lower mass planets
are even more common. With more recent data, Shvartzvald et al. (2016)
found that $\sim$ 55\% of these low mass (K-M dwarf) stars hosted planets,
with Neptune-mass planets being about 10 times as common as Jupiter-mass
planets. Sumi et al. (2011) interpreted short-time-scale microlensing events
(less than 2 days) as evidence for a population of unbound or distant
Jupiter-mass objects that are about twice as common as main-sequence
stars. Sumi et al. (2011) argued that these objects were most likely
to be unbound gas giants that were ejected from protoplanetary disks,
but Veras \& Raymond (2012) found that planet-planet scattering
could not be expected to form that many objects. Their analysis
was restricted to gas giants formed by core accretion, i.e., their
planets were initially located at a few AU from their stars. Ma et al. (2016) 
similarly found that core accretion could not produce the number of ejected 
planets implied by the Sumi et al. (2011) observations. Forgan et al. (2015) 
and Vorobyov (2016), on the other hand, propose that disk fragmentation 
in the outer disk regions will lead to the frequent ejection of brown 
dwarfs and gas giants, perhaps explaining the Sumi et al. (2011) results.
Even if these objects are bound, their orbits must lie beyond 10 AU
from their stars, a region where disk instability is likely to be the 
main formation mechanism, rather than core accretion (see below). Contrary
to the assertions by Sumi et al. (2011), direct imaging surveys searching
for such distant but bound gas giants do not exclude the microlensing
population from being bound objects (Quanz et al. 2012).
Either way, gravitational instability seems to be implicated in explaining
the Sumi et al. (2011) gas giant population.

\subsection{Directly Imaged Exoplanets on Wide Orbits}

 The discovery of the HR8799 bcde exoplanet system (Marois et al. 2008, 2010)
is as remarkable in its own way as the millimeter-wave image of the HL Tau disk 
(ALMA Partnership et al. 2015). This system consists of {\it four} gas giant 
planets with masses of about 5 to 7 $M_J$, orbiting at distances of about 
14, 24, 38, and 68 AU from an A5 star of 1.5 $M_\odot$ with an 
age of $\sim$ 30 Myr. 51 Eridani is a younger ($\sim$ 20 Myr) F0IV star with 
a mass of $1.75 \pm 0.05 M_\odot$ and with a $\sim 2 M_J$ planet orbiting 
at a distance of $\sim 13$ AU (Macintosh et al. 2015). HD 100546 
is an even younger ($\sim$ 10 Myr) Herbig Ae/Be star with a mass of 
$2.4 \pm 0.1 M_\odot$, which has a gas giant planet orbiting at a distance of 
$\sim 53$ AU, and a possible second planet at $\sim$ 14 AU (Quanz et al. 2015).
Finally, LkCa 15 is an extremely young ($\sim$ 2 Myr) solar-type
star with two likely protoplanets orbiting $\sim$ 15 and 19 AU away,
with planet masses no greater than $\sim 5-10 M_J$ being required
for orbital stability (Sallum et al. 2015).

 The M0 close binary system ROXs 42B is orbited at $\sim$ 150 AU by
a third body with a mass in the range of 6 to 15 $M_J$ (Currie et al. 2014).
This third object appears to blur the line between exoplanets (with masses less
than $\sim 13 M_J$), such as the HR 8799 bcde planetary system  
(Marois et al. 2008; 2010), and brown dwarfs. It also therefore blurs the
line between the formation mechanisms of wide and close binary stars,
brown dwarfs, and possibly of gas giant exoplanets, namely protostellar and 
protoplanetary disk fragmentation (Currie et al. 2014). As one example,
Konopacky et al. (2016) found a brown dwarf companion with a mass of 
$\sim 30 M_J$ orbiting 20 AU from HR 2562, a 1.3 $M_\odot$ star 
with a debris disk. Such brown dwarfs are 
widely believed to form by disk fragmentation, based on both observational
(e.g., Ma \& Ge 2014) and theoretical (e.g., Li et al. 2015) evidence.

\subsection{Giant Planet Formation on Wide Orbits}

 Given the observational evidence for a growing number of extrasolar giant 
planets orbiting at distances well beyond the orbits of Jupiter and Saturn,
the question arises as to their formation mechanism. The formation of 
the solar system's outer giant planets, Uranus and Neptune, is considered
problematical for the core accretion (CA) mechanism. Levison et al. (2015)
claimed to have solved the problem of forming all four of the solar 
system's giant planets by relying on the pebble accretion mechanism 
(e.g., Bitsch et al. 2015) to grow giant 
planet cores rapidly, before the solar nebula gas disappeared. Their
CA models led to the formation of 1 to 4 giant planets between 5 and 15 AU, 
as desired. However, their initial conditions consisted of a protoplanetary
disk that is assumed to be passive, with a surface density profile (Figure 1)
that is nearly identical to that of the gravitationally unstable disk 
models studied by Boss (2013, 2015) and by this paper. Such a moderately 
massive disk will not be passive, however, and 
the resulting spiral arms will play havoc with the orbits of the solid bodies
that are trying to grow by pebble accretion, as shown by Boss (2013, 2015).
As a result, the formation of even the solar system's outer giant planets should 
be considered to remain problematical. Bromley \& Kenyon (2016) studied
CA models in disks as massive as 0.5 $M_\odot$ in order to try to 
explain the possible existence of a massive, ninth planet orbiting
beyond 100 AU, yet another challenge for solar system formation models.

 Chambers (2006) studied the CA process in a disk with a radius of 50 AU,
and found that no giant planets more massive than $\sim 0.2 M_J$ were
formed beyond $\sim$ 8 AU. Kenyon \& Bromley (2015) found that in a
suitably massive disk, CA could lead to the formation of an icy 
super-Earth at distances of 125 to 250 AU, but that this would require
1 to 3 Gyr, roughly a factor of $10^3$ times too slow to allow such an 
icy core to accrete a gaseous envelope and form a gas giant.
Chambers (2016) studied CA in the pebble accretion scenario for disks
with radii of 50 to 100 AU, finding that gas giant planet formation
was limited to orbital distances of $\sim$ 15 AU or less. A similar 
result was found in the CA models of Coleman \& Nelson (2016). At best, CA 
would seem to require the formation of massive icy cores in the inner
disk, which are then scattered to the outer disk by close encounters
with one or more inner gas giants. Once in the outer disk, the icy cores 
could accrete outer disk gas, which could circularize their initially
highly eccentric orbits (Kikuchi et al. 2014). However, objects
formed by this mechanism are predicted to be extremely rare, 
occurring in only $\sim$ 0.1\% of the cases studied by population
synthesis models (Kikuchi et al. 2014). Such a scattering process is also
likely to be hazardous to the health of inner terrestrial planets
(Kaib \& Chambers 2016).

 On the other hand, Boss (2011) found that disk instability (DI) could form 
numerous gas giants with initial masses in the range of 1 to 5 $M_J$, initial 
distances of 30 to 70 AU, and initial eccentricities of $\sim$ 0 to 0.35,
around solar-type protostars with masses from 0.1 to 2 $M_\odot$.
The initial disk mass was assumed to scale with the protostellar mass, 
with the result that the more massive stars had more giant exoplanets,
as expected. Forming the HR 8799 exoplanet system thus seems quite
possible thorough the DI mechanism (Boss 2011). In fact, 
forming gas giants at large distances by disk instability has now
become a part of the conventional wisdom of planet formation theory 
(e.g., Boley 2009; Boley \& Durisen 2010; Boley et al. 2010; 
Nero \& Bjorkman 2009; Meru \& Bate 2010; Kratter \& Murray-Clay 
2011; Rogers \& Wadsley 2012; Vorobyov et al. 2013;
Madhusudhan et al. 2014; Rice et al. 2015; Young \& Clarke 2016).
Meru (2015) found that fragments formed by DI in the outer disk
could even trigger the formation of more fragments in the inner disk.

 Galvagni et al. (2012) found that clumps formed by DI in the outer disk at 
$\sim$ 100 AU could contract fast enough ($\sim 10^3$ yr) to begin rapid
dynamic collapse without being tidally disrupted. Nayakshin (2015a,b)
showed that pebble accretion by a gaseous protoplanet could accelerate
its collapse by changing its opacity, again helping avoid tidal disruption.
Boss (2013) found that giant planets orbiting in a marginally gravitationally
unstable disk did not undergo monotonic orbital migration, but rather
underwent random inward and outward excursions, driven by the gravitational
actions of the spiral arms, for several thousand years or more.
Stamatellos (2015) found that a giant planet formed in the outer disk would 
soon open a disk gap, preventing its migration to the inner disk (cf., Vorobyov 
2013), and that it would not accrete sufficient gas to become a brown dwarf,
rather than a gas giant planet (cf., Kratter et al. 2010).

\section{Hydrodynamical Models of Disk Instability}

 Given this strong observational evidence that a mechanism similar
to disk instability must be able to form at least some exoplanets,
it behooves us to consider the current status of disk instability
theoretical modeling efforts. We first note the present status of
the more computationally challenging models that involve 3D radiative
transfer, and then introduce the computationally simpler $\beta$ cooling
approach.

\subsection{Radiative Transfer Models}

 Boss (1997) began the detailed study of DI with 3D models that 
used simplified thermodynamics for the gaseous disk, namely isothermal 
or adiabatic gas pressure laws. Boss (2001) then calculated the first 
DI models with 3D radiative transfer (i.e., radiation hydrodynamics - RHD), 
finding that disk fragmentation was still possible. Boley et al. (2006)
found in their 3D RHD models that the clumps that formed did not last long.
Durisen et al. (2007) summarized much of the detailed work on 3D radiative
transfer models of the DI mechanism, both from the side of models where
clump formation occurred (e.g., Boss 2001, 2002, 2007; Mayer et al.
2007) and from models where clump formation was less robust (e.g., Boley et al. 
2006, 2007). While many numerical factors come to play in these calculations,
e.g., grid resolutions for finite difference (FD) codes and smoothing lengths
for smoothed-particle hydrodynamics (SPH) codes (e.g., Mayer et al. 2007), 
flux limiters for radiative transfer in the diffusion approximation, and  
accuracy of the gravitational potential solver, the key issue was determined 
to be whether a protoplanetary disk could remain sufficiently cold for
spiral arms to collide and form self-gravitating clumps that could
contract toward planetary densities without re-expanding, being disrupted
by tidal forces from the central protostar, or meeting some other equally
unkind fate, i.e., the radiative transfer solver. Boley et al. (2006) showed
that their cylindrical coordinate RHD code could reproduce an analytical 
radiative transfer solution, while Boss (2009) demonstrated the agreement 
of his spherical coordinate RHD code on two other analytical radiative transfer 
solutions. Helled et al. (2014) updated the situation in their review paper 
without reaching a definitive conclusion regarding the validity of the DI 
hypothesis in RHD models (e.g., the work by Boss 2008, 2009, 2010, 2011, 2012; 
Boley \& Durisen 2008, 2010; Boley 2009; Cai et al. 2010; Meru \& Bate 2010). 
More recently, Steiman-Cameron et al. (2013) studied the effect of spatial 
resolution on cooling times in RHD models, finding convergence for optically 
thick, inner regions, but not for optically thin, outer regions.
Tsukamoto et al. (2015) used 3D RHD models to 
follow the formation of disks, starting from collapsing molecular cloud 
cores, finding that radiative heating from the interstellar medium 
could have a significant effect on the fragmentation process.

\subsection{Beta Cooling Models}

 Gammie (2001) suggested that the outcome of gravitational instability
depended on the beta parameter $\beta = t_{cool} \Omega$,
where $t_{cool}$ is the disk cooling time and $\Omega$ is the
local angular velocity. Gammie (2001) found that for $\beta > 3$,
the disk would become gravitoturbulent with a Toomre $Q$ value $\sim 1$,
whereas for $\beta < 3$, the disk would fragment. 
Lodato \& Rice (2004) and Mejia et al. (2005) also discussed
the importance of cooling times for disk fragmentation.
The use of the critical 
value of $\beta = 3$ to predict the fragmentation of protoplanetary 
disks was first called into question by the 3D SPH hydrodynamical
models of Meru \& Bate (2011a,b), who found that when 
sufficiently high spatial resolution was employed, even disks previously 
thought to be stable underwent fragmentation into clumps. 
Meru \& Bate (2011b) even suggested that in the absence of clear
indications of numerical convergence, a critical value might not exist.
Meru \& Bate (2012) found evidence for convergence with both 3D SPH
and 2D FD (FARGO) codes on a critical value of $\beta > 20$, and perhaps
as large as $\sim$ 30. 

 Gammie (2001) had studied two dimensional (2D), shearing-sheet disk models 
with a maximum numerical resolution of $1024^2$. Baehr \& Klahr (2015) 
also studied 2D shearing sheet models, but with a refined cooling law, 
and found that with their highest resolution models ($4096^2$), the 2D 
disks fragmented even for $\beta = 10$, but not when the resolution 
was $2048^2$ with $\beta = 10$. Paardekooper (2012) studied 2D
shearing sheet models with the FARGO FD code, finding fragmentation to occur
for $\beta$ values as high as 20, provided the integration was followed
long enough in time. 

 Rice et al. (2014), however, argued that numerical
problems involving the implementation of the disk cooling in SPH
codes and the handling of artificial viscosity in both SPH and the 
FARGO FD codes may have led to higher critical $\beta$ values.
They claimed that their SPH simulations converged on a critical
value of $\beta$ between 6 and 8.
However, Zhu et al. (2015) have shown that SPH models will not converge
to the continuum limit if only the total number of particles ($N$) is
increased to $\infty$; formal numerical convergence is only possible
when $N \rightarrow \infty$, the smoothing length $h \rightarrow 0$,
and the number of neighbor particles within the smoothing volume
$N_{nb} \rightarrow \infty$. The latter constraint is seldom applied,
lending suspicion to the claims of numerical convergence of many previous
SPH studies. 

 Young \& Clarke (2015) used both 2D SPH and 2D finite difference models to 
suggest two possible pathways to fragmentation: dynamic collapse when 
$\beta < 3$, or quasi-static contraction when $\beta < 12$, for models with 
effectively $2048^2$ resolutions. Evidently even for simplified $\beta$ 
cooling prescriptions, enough numerical questions and differences remain that 
the critical $\beta$ value for fragmentation, if there is one, cannot be 
constrained better than to lying in the range of $\sim$ 10 to $\sim$ 30.

\subsection{Initial Toomre $Q$ Values}

 Takahashi et al. (2016) pointed out the crucial importance of the Toomre $Q$
value: fragmentation occurs when the Toomre $Q$ value drops to $\sim$ 0.6 
inside the spiral arms, regardless of the value of $\beta$, and that low
$\beta$ values do not necessarily result in fragmentation.
Gammie (2001) studied 2D disks with $Q = 1$ initially. The Paardekooper 
(2012) and Baehr \& Klahr (2015) 2D models also started with $Q = 1$ 
throughout the disk. Young \& Clarke (2015) started with disks with 
$Q \approx 1$. Meru \& Bate (2012) and Rice et al. (2014) started with disks
with a minimum of $Q = 2$ at the outer disk edge.
Rafikov (2015) has argued that gravitoturbulent disks without rapid
cooling should settle into a quasi-stationary, fluctuating state
where $Q$ remains close to a constant value $Q_0 \sim 1$. 
The concept of a quasi-steady balance of disk heating and cooling
dates back to Pringle (1981), and has been investigated in many models
by the Indiana University group (e.g., Steiman-Cameron et al. 2013), 
yielding $Q \sim 1.5$ to 2. This balance has been suggested to be an 
alternative to fragmentation (e.g., Durisen et al. 2007).

 Backus \& Quinn (2016) found that the critical initial value of $Q$ for fragmentation
depended on how stable the initial disk equilibrium model was: less stable 
initial disk equilibria fragmented for higher $Q$ values.
Evidently initial $Q$ values in the range of 1 to 2 are considered
reasonable starting values for DI models with $\beta$ cooling.

\section{Numerical Methods}

 In order to allow a direct comparison with previous work, the new models 
were calculated with the same basic code that has been used in all of 
the author's previous studies of disk instability. The primary change
was to replace the radiative transfer subroutine that calculates the
energy changes due to radiative flux with the $\beta$ cooling formula.
The numerical code solves the three dimensional equations of hydrodynamics 
and the Poisson equation for the gravitational potential and is 
second-order-accurate in both space and time. A complete description
of the entire code is given by Boss \& Myhill (1992), while the updated 
energy equation of state is described by Boss (2007). Also, the central 
protostar is effectively forced to wobble in order to preserve the 
location of the center of mass of the entire system 
(Boss 1998, 2012), which is accomplished by altering the apparent 
location of the point mass source of the star's gravitational potential 
in order to balance the center of mass of the disk. 
Central massive sink cell particles behave in
exactly this same way to preserve the center of mass of the system
when used in, for example, the FLASH hydrodynamics code.

 As usual, explicit artificial viscosity (AV) is not used in the 
models. Boss (2006) found that small amounts of tensor AV had little effect 
on fragmentation, while large amounts could suppress fragmentation, as
was also found by Pickett et al. (2000). AV is designed to stabilize and
capture the microphysics of heating in strong shocks, and the effect
of its inclusion in $\beta$ cooling models remains to be investigated.

 The numerical code solves the specific internal energy $E$ equation 
(Boss \& Myhill 1992): 

$${\partial (\rho E) \over \partial t} + \nabla \cdot (\rho E {\bf v}) =
- p \nabla \cdot {\bf v} + L, $$

\noindent
where $\rho$ is the gas density, $t$ is time, ${\bf v}$ is the velocity, 
$p$ is the gas pressure, and $L$ is the time rate of change of energy 
per unit volume, which is normally taken to be that due to the transfer 
of energy by radiation in the diffusion approximation. Here we define $L$
in terms of the $\beta$ cooling formula (Gammie 2001), as follows. With
$\beta = t_{cool} \Omega$, $t_{cool}$ is defined as the ratio of the specific
internal energy to the time rate of change of the specific internal
energy. We thus define $L$ to be:

$$L = - {\rho E \Omega \over \beta}, $$

\noindent
where $\Omega$ is the local angular velocity of the gas in prograde
rotation ($\Omega > 0$). Evidently
$L$ is always negative with this formulation, i.e., only cooling
is permitted. 

 It is important to note that in spite of the $\beta$ cooling
formula, the disk temperature at a given radial distance from the
central protostar is not allowed to fall below its initial value.
This means that initially high $Q$ disks cannot become more 
gravitationally unstable solely due to cooling to lower temperatures
than the initial state, regardless of the cooling parameter $\beta$,
and can only become gravitationally unstable by transporting disk
mass such that the local disk surface density increases, thereby
lowering $Q$ locally. This assumption is critical to understanding
the results to be presented for initially high $Q$ disk models.
This minimum temperature constraint was also imposed in all of the 
previous disk instability models in this series, and so was retained
here in order to allow a direct comparison with those earlier models.
A future paper will address the outcome of models similar to these, but
where the disk temperature is allowed to cool below the initial value. 
An alternative justification for this temperature constraint is to 
consider the case of disks being heated by protostellar irradiation, e.g.,
Cai et al. (2008) and Takahashi et al. (2016), in which case disks
might become hot enough to suppress fragmentation altogether.

 The equations are solved on a spherical coordinate grid with $N_r = 100$
or 200 radial grid points (as well as the central grid cell, which 
contains the central protostar), $N_\theta = 23$ theta grid points,
distributed from $\pi/2 \ge \theta \ge 0$, and $N_\phi = 512$ or 1024
azimuthal grid points. $N_r$ and $N_\phi$ are increased to their higher values
once the disk forms dense spiral arms and the Jeans and Toomre length
criteria begin to be violated (see below). The radial grid is uniformly 
spaced with either $\Delta r = 0.16$ AU or 0.08 AU, for $N_r = 101$
or 200, respectively. The radial grid extends from 4 to 20 AU.
The mass of disk gas flowing inside 4 AU is added to the central 
protostar, whereas that reaching the outermost shell at 20 AU loses
its outward radial momentum but remains on the active hydrodynamical grid.
The $\theta$ grid points are compressed into the midplane to ensure 
adequate vertical resolution ($\Delta \theta = 0.3^o$ at the midplane). The 
$\phi$ grid is uniformly spaced in $2 \pi$. The number of terms in the 
spherical harmonic expansion for the gravitational potential of the 
disk is $N_{Ylm} = 48$ for all phases of evolution. 

 The numerical resolution is increased during the evolutions in order
to avoid violating the Jeans length (e.g., Boss et al. 2000) and Toomre
length criteria (Nelson 2006). Both criteria are monitored throughout
the evolutions (e.g., Boss 2011) to ensure that any clumps that might 
form are not numerical artifacts. The Jeans length criterion consists of
requiring that all of the grid spacings in the spherical coordinate
grid remain smaller than 1/4 of the Jeans length $\lambda_J =
\sqrt{ \pi c_s^2 \over G \rho}$, where $c_s$ is the local sound
speed, and $G$ the gravitational constant.
Similarly, the Toomre length criterion consists of requiring that 
all of the grid spacings remain smaller than 1/4 of the Toomre 
length $\lambda_T = (2 c_s^2 / G \Sigma)$, where $\Sigma$ is the 
mass surface density. When one of these two criteria is violated,
the calculation is halted, and the grid spatial resolution is doubled
in either the radial or azimuthal direction, as necessary, by
dividing each cell into half in the relevant direction. Mass and
momentum are conserved by this process.

 Once well-defined clumps form, the Jeans and Toomre length
criteria will eventually again be violated at the maximum densities of 
the clumps, even with the $N_r = 200$ and $N_\phi = 1024$ grids.
At this point, the cell with the maximum clump density is drained of 90\% 
of its mass and momentum, which is then used to form a virtual protoplanet
(VP) initially at the cell center (e.g., Boss 2005, 2013). The VPs 
thereafter orbit around the disk, subject to the gravitational forces of 
the disk gas and the central protostar, as well as those of any other
VPs. The disk gas is similarly subject to the gravity of the VPs. VPs that
orbit to the inner boundary at 4 AU or to the outer boundary at 20 AU 
are simply removed from the remainder of the evolutions. The VPs are
allowed to gain mass $\dot M$ at the rate (Boss 2005, 2013) given by the 
Bondi-Hoyle-Lyttleton (BHL) formula (e.g., Ruffert \& Arnett 1994):

$$ \dot M = { f 4 \pi \rho (G M)^2 \over (v^2 + c_s^2)^{3/2}}, $$

\noindent
where $f$ is a dimensionless coefficient, $M$ is the VP mass, 
and $v$ is the speed of the VP respect to the disk gas. 
The VPs also accrete orbital angular momentum from the disk gas,
by accreting an amount of momentum from the local hydrodynamical
cell proportional to the mass being accreted from that cell 
in such a way as to guarantee the conservation
of the total orbital angular momentum. This insertion of VPs in
regions where the Jeans and Toomre length criteria can no longer
be met is identical to the manner in which {\it sink particles} are inserted
into adaptive mesh refinement codes, such as FLASH, when no further
sub grids are allowed to be formed (e.g., Federrath et al. 2010;
Klassen et al. 2014).

 These models used $f = 1$ in order to maximize the gas accretion
by the VPs and to minimize subsequent Jeans or Toomre length violations.
If the Jeans or Toomre criterion is nevertheless subsequently violated in 
the immediate vicinity (10 cells) of an existing VP, the violation is 
ignored, as the VP is expected to accrete or disrupt this high 
density gas. Violations more than 10 cells away result in the
creation of a new VP. Along with the maximum spatial resolution required,
the number of VPs formed thus gives a basic estimate of the extent 
to which a given set of initial conditions leads to a gravitationally
unstable disk capable of fragmenting into clumps. 

\section{Initial Conditions}

 The models all begin from the same initial disk model as that used
by Boss (2001, 2005, 2007, 2008, 2012, 2013), differing only in the minimum
initial Toomre $Q$ value and the $\beta$ parameter. The initial 
density is that of an adiabatic, self-gravitating, thick disk in 
near-Keplerian rotation about a stellar mass $M_s$ (Boss 1993):

$$ \rho(R,Z)^{\gamma-1} = \rho_o(R)^{\gamma-1} - \biggl( 
{ \gamma - 1 \over \gamma } \biggr) \biggl[
\biggl( { 2 \pi G \sigma(R) \over K } \biggr) Z +
{ G M_s \over K } \biggl( { 1 \over R } - { 1 \over (R^2 + Z^2)^{1/2} }
\biggr ) \biggr], $$

\noindent where $R$ and $Z$ are cylindrical coordinates, $\sigma(R)$ is 
the surface density, $K = 1.7 \times 10^{17}$ (cgs units) and $\gamma = 5/3$.
The initial midplane density ensures near-Keplerian rotation throughout the 
disk:

$$\rho_o(R) = \rho_{o4} \biggl( {R_4 \over R} \biggr)^{3/2}, $$

\noindent 
where $\rho_{o4} = 10^{-10}$ g cm$^{-3}$ and $R_4 = 4$ AU, the inner edge 
of the numerical grid. The resulting initial disk mass is 0.091 $M_\odot$,
while the initial protostellar mass is $M_s = 1.0 M_\odot$. 
Note that this initial approximate equilibrium disk state does
not depend explicitly on the outer disk temperature, and was chosen
to be a state dominated by rotational support rather than by gas pressure,
at least in the radial direction (Boss 1993). As a result, this same
initial density distribution is used for all the models, regardless of
the variations in the outer disk temperatures.

 Table 1 lists the initial conditions chosen for the models. The initial 
outer disk temperatures $T_o$ are varied in order to alter 
the initial minimum values of the Toomre (1964) $Q$ gravitational 
stability parameter, decreasing monotonically outward from 
highly stable $Q_{max} = 9$ values at 4 AU to marginally stable 
$Q_{min}$ values ranging from 1.3 to 2.7 beyond 10 AU. Values of $\beta$
of 1, 3, 10, 20, 30, 40, 50, and 100 were used in the models. These 
choices for the initial disk conditions, $Q_{min}$, and $\beta$ are consistent
with the observational constraints and with the papers on $\beta$ 
cooling models that were cited previously.

 Figure 1 shows the initial surface density profile for the models,
compared to the power law assumed in the core accretion models by 
Levison et al. (2015). We shall see that the passive disk assumed in these
core accretion models is likely to be gravitationally unstable for
realistic disk midplane temperatures. Figure 2 displays the initial Toomre
(1964) $Q$ profiles for three representative sets of models, those with
outer disk temperatures of 40 K, 90 K, and 180 K, leading to initial 
minimum $Q$ values beyond 10 AU of 1.3, 2.0, and 2.7, respectively.
Midplane temperatures at the inner disk edge at 4 AU are 600 K for all
models, while the outer disk temperatures are varied as listed in Table 1.
The initial midplane temperature profiles are based on the 2D RHD models
by Boss (1996), as used by Boss (2001, 2005, 2007, 2008, 2012). 

 Figure 3 displays the initial midplane density distribution for all
of the models, along with the initial midplane temperature distribution
for the models with initial minimum Toomre $Q_i = 1.3$, i.e., the models
with outer disk temperatures of 40 K. While the initial temperature
distributions are perfectly axisymmetric about the rotation axis (center
of each plot), the initial density distributions are non-axisymmetric
at a low level: random cell-to-cell noise at the level of 1\% has
been added to the density distribution, along with $m = 1, 2, 3,$ and 4
modes of amplitude 0.01, as in previous models (e.g., Boss 1998).

\section{Results}
 
  Table 1 lists the basic results for all of the models, namely
the final times reached, the final spatial resolution employed, and 
the maximum number of virtual planets ($N_{VP}$) that formed and 
existed simultaneously, usually close to the final number of VPs.
The final times reached ranged from 244 yrs to 978 yrs. For comparison, 
the initial orbital period of the disk at the inner 
edge (4 AU) was 8.0 yr, whereas the initial orbital period at the 
outer edge (20 AU) was 91 yrs. The key point is that all the models
were evolved far enough in time that there was no evidence for any 
further significant growth of non-axisymmetry, i.e., no need to further
refine the spatial grid. Some models still required the creation
of additional VPs, but this was balanced by the loss of VPs that hit
either the inner or outer grid boundaries, indicating that a semblance
of a steady state configuration had been reached.
Because the Boss \& Myhill (1992) code employed has not been parallelized, 
each model was computed continuously on a single, dedicated cluster core.
Most models were run for as long as 2.3 yrs on the DTM flash cluster,
while others were run for as long as 7 months on the considerably faster 
Carnegie memex cluster at Stanford University.

 Figure 4 summarizes the results from Table 1 as follows. Models where
the initial spatial resolution of $N_r = 100$ and $N_\phi = 512$ sufficed
for the entire evolution without violating the Jeans or Toomre
constraints experienced little growth of non-axisymmetry (red dots),
while those that needed to be refined to $N_r = 200$ with $N_\phi = 512$
experienced moderate growth (green dots), and those that ended with
$N_r = 200$ and $N_\phi = 1024$ experienced significant growth of
spiral arms (blue dots). The remaining models continued to violate
the Jeans or Toomre criteria, even with $N_r = 200$ and $N_\phi = 1024$,
necessitating the creations of VPs at the location of the violations. 
The black symbols in Figure 4 represent the maximum number of VPs
created for those models, ranging from a black circle for a single VP,
to a bar for two VPs, to a three-pointed star for three VPs, etc.

 Figure 4 demonstrates that the assumed initial conditions are just
as important for determining the success of a possible disk gravitational
instability in forming self-gravitating clumps as is the choice of the
$\beta$ cooling parameter. All of the models with $Q_i = 1.3$ fragmented 
and formed from 4 to 10 VPs, regardless of the cooling rates 
investigated, from $\beta$ = 1 to $\beta$ = 100. This somewhat surprising
result means that if a disk is initially extremely gravitationally
unstable, self-gravitating clumps can form rapidly without being
stifled by compressional heating during their assembly. 
However, the time at which the first VP formed depended strongly on 
the value of $\beta$, with the first VP in models 1.3-1 and 1.3-3 forming 
after 104 yrs and 118 yrs, respectively, compared to forming only 
after 190 yrs and 212 yrs, respectively, for models 1.3-50 and 1.3-100.
Clearly even for an initially highly unstable disk, the cooling
rate affects the time evolution of the fragmentation process. 
At the same 
time, the models starting with $Q_i = 2.7$ show that such an initially
gravitationally stable disk cannot become non-axisymmetric enough to
undergo fragmentation, even for $\beta = 1$. 
As noted previously, this
result is largely a result of the constraint that the disk temperatures
cannot drop below their initial values, in spite of vigorous cooling,
and can only become more gravitationally unstable by increasing the 
disk surface density in a limited region of the disk, e.g., by forming a
ring.
For initial Toomre $Q$ 
values in between these two extremes, the general result is a smooth
progression toward more of a tendency toward fragmentation at a fixed 
value of $\beta$ (e.g., for $\beta$ = 3 and 10 as $Q_i$ is lowered). 
However, the transition
from stable to unstable with regard to forming VPs is not necessarily
completely monotonically dependent on $Q_i$ (i.e., for $\beta$ = 1 and 
100, where as $Q_i$ is decreased, successive models can fragment and then
resist fragmentation, before eventually fragmenting again). This result
demonstrates the stochastic nature of gravitationally unstable disks,
where the spiral arms that form repeatedly and interact with each
other may or may not happen to combine in a constructive wave sense 
and form a clump dense enough to require the creation of a VP.
Given this ambiguity, the critical value for fragmentation appears
to be $\beta_c \sim 1$ for $Q_i \sim 2.2$, 
$\beta_c \sim 10$ for $Q_i \sim 1.9$, and
$\beta_c \sim 50$ for $Q_i \sim 1.6$, roughly speaking.

 Figure 5 presents the final midplane density distributions for
the four models at the extremes of the parameter space investigated,
namely models 1.3-1, 1.3-100, 2.7-1, and 2.7-100. The two models
with $Q_i$ = 1.3 both clearly formed numerous spiral arms and
clumps, regardless of whether $\beta$ = 1 or 100, though in the 
latter case the spiral arms are broader and nowhere near as sharply 
defined. Neither of the two models with $Q_i$ = 2.7 formed any
significant non-axisymmetric features, and the model with
$\beta$ = 100 looks the most similar to the initial disk model
(Figure 3a), in spite of having evolved for 414 yrs. Figure
6 displays the corresponding midplane temperature distributions for 
these same four end-member models. The slower cooling rate in model 
1.3-100 (Figure 6b) compared to that of 1.3-1 results in well-defined,
moderately warm spiral arms throughout the disk, whereas the rapid
cooling assumed in model 1.3-1 (Figure 6a) allows the disk to cool
back down to its initial temperature throughout most of the disk.
For the high $Q_i$ models, Figure 6c shows that model 2.7-1
maintains a disk temperature essentially unchanged from its
initial conditions, as a result of the rapid cooling, whereas
in Figure 6d, model 2.7-100 shows several distinct rings of somewhat
hotter gas than the initial temperature distribution.

 Figure 7 shows the location of the nine VPs still active in
model 1.3-1 at the end of the evolution (224 yrs), the same
time as the midplane density contours shown in Figure 5a. It is
evident that the nine VPs are not aligned with the numerous
spiral arm features seen in Figure 5a, as is to be expected,
given that the VPs begin their existence with an initial mass
of the order of a fraction of a Jupiter mass, and as such are
not subject to the tendency for gas drag to force small 
(e.g., cm-size) particles to remain in the vicinity of spiral
arms (e.g., Boss 2013, 2015).

 The time evolution of the ensemble of VPs formed in the eight
models with $Q_i$ = 1.3 is presented in Figure 8, sampled
roughly every 40,000 time steps. Starting with masses only
a small fraction of a Jupiter mass, the VPs accrete mass at the
BLH rate and soon reach masses as high as 5 $M_J$. The VPs
tend to be created in the sweet spot between 6 AU and 10 AU, 
where the disk is cool enough and massive enough to support the
growth of strong spiral arms. Once formed there, the VPs can migrate
through disk interactions (e.g., Boss 2013) both inwards and
outwards, with significant numbers hitting either the inner or
outer disk boundaries and thereafter being removed from the 
calculations.

 Finally, Figure 9 compares the VPs formed in the $Q_i$ = 1.3
models with the presently known distribution in mass and semi-major
axis of exoplanets between 4 AU and 20 AU. While not intended
to be a rigorous exoplanet population synthesis on a par with
those presented in Section 2.5, Figure 9 hints that the disk
instability mechanism may be able to crudely match the exoplanet 
demographics for gas giants with semi-major axes of 4 AU to 6 AU,
and if so, this approach would predict an as yet mostly undetected
significant population of Jupiter-mass gas giants orbiting at
distances of about 6 AU to 16 AU. Refining the implications of
these disk instability models for explaining the known exoplanet
demographics, and for predicting what more might be discovered by
future exoplanet surveys (e.g., by microlensing and direct imaging
searches with the NASA WFIRST mission) is a promising subject for 
further work.

\section{Discussion}

 These models were intended in part to provide a comparison 
between the disk stability models previously presented in this series 
(e.g., Boss 2001, 2007, 2008, 2012), which employed diffusion approximation
radiative transfer, with new models using the $\beta$ cooling approach.
The previous models all explored variations on numerical parameters
such as spatial grid resolution, equations of state, temperature
assumptions, and diffusion approximation flux-limiters, with the 
result that the disks, with initial $Q_{min} = 1.3$, nevertheless
fragmented into clumps. Given the results found here for the $Q_{min} = 1.3$ 
initial disks, which all formed numerous VPs, even with $\beta$ as
high as 100, the basic results of the previous diffusion approximation 
models starting from these initial conditions do not appear to be as 
contentious as they have at times seemed (e.g., Durisen et al. 2007; 
Helled et al. 2014).

 Section 2.3 noted that several indicators of protoplanetary disk 
temperatures at distances of $\sim$ 10 to 20 AU imply midplane
temperatures of $\sim$ 40 to 70 K, i.e., values of $Q_i$ from
about 1.3 to 1.7 for the present disk models. Figure 4 shows that
such disks are unstable to fragmentation on short time scales,
regardless of the assumed value of $\beta$. Figure 1 shows that
these disk models have an initial surface density profile quite
similar to that assumed in the Levison et al. (2015) models of
solar system formation by pebble accretion. Evidently the background
passive gas disk adopted for the Levison et al. (2015) efforts is 
unlikely to be passive, calling into question their results regarding
the formation of our giant planets.

 Boss (2004) found evidence in 3D diffusion approximation models of
disk instability for convective upwellings that were able to cool the 
outer disk at a rate equivalent to $\beta \sim 6$, leading to 
fragmentation, as expected based on the results shown in Figure 4
[Note, however, that Boley \& Durisen (2006) and Lyra et al. (2016)
have interpreted such upwellings as hydraulic jumps, rather than as the
result of convection).]
As previously noted, Meru \& Bate (2011b) suggested that a critical value 
of $\beta$ for fragmentation might not exist, a suspicion that is supported
by the weak dependence of the results shown in Figure 4 on $\beta$
for a fixed value of $Q_i$, i.e., the initial conditions assumed
are more important in general than the choice of $\beta$.
Takahashi et al. (2016) found a similar result to that seen in
Figure 4, namely that fragmentation can occur whenever $Q$ becomes
sufficiently small, regardless of $\beta$, and that low $\beta$
alone does not guarantee fragmentation. In particular, Takahashi 
et al. (2016) found that fragmentation would occur when $Q$ dropped
to 0.6 in a spiral arm. Figure 10 displays the radial profile of
$Q$ in model 1.3-1 at the time (104 yrs) when a VP needed to
be inserted into the calculation: it can be seen that the minimum
azimuthally averaged value of $Q$ at that time was 0.61, seemingly
in remarkable agreement with the results of Takahashi et al. (2016).
However, for the seven other models with $Q_i = 1.3$, the equivalent minimum
$Q$ value ranges from 0.69 to 1.0. Given that these are azimuthally
averaged $Q$ values, however, the basic agreement with Takahashi et al. 
(2016) holds. In fact, the first VP in model 1.3-1 was inserted 
at 7.2 AU, the radius of the innermost local minimum in the azimuthally 
averaged $Q$ seen in Figure 10. The $Q$ value of the actual midplane grid 
cell where this first VP was inserted was $\sim 0.1$, considerably lower
still.

\section{Conclusions}

 The present set of models has shown that the disk instability
mechanism for gas giant planet formation depends even more strongly
on the initial conditions assumed in the models than on the assumed
disk cooling rate $\beta$, based on a suite of 3D models with 
identical handling of the spatial resolution of the grid. Hence the
evolution of protoplanetary disks into gravitationally unstable
configurations is as important a factor to consider as the detailed
heating and cooling processes in the disks. The models imply that
if a significant fraction of protoplanetary disks can form that are
similar to those assumed here, then there should be an equally significant 
population of gas giants remaining to be discovered with separations 
of order 15 AU from solar-type stars. The NASA WFIRST mission may
very well test this prediction, based on a combination of exoplanet
searches by gravitational microlensing and by coronagraphic direct 
imaging. 

 In order to help determine what is the proper value of $\beta$ to use in
disk instability models, the author is currently running a suite of
eight flux-limited diffusion approximation (FLDA)
models similar to the present suite, with initial minimum $Q$ values
ranging from 1.3 to 2.7, and with the same approach of adding spatial
resolution and VPs when demanded by the Jeans and Toomre criteria,
with the goal of comparing these results for varied beta cooling models
with those of FLDA models, for the same spatial resolution and same
initial conditions (i.e., initial Toomre $Q$). Previous FLDA disk
instability models by Boss (2008, 2012) used different spatial
resolutions than the present models and so cannot be used for a valid
comparison. However, the suite of models currently running will allow
the proper value of $\beta$ that should be used in disk instability
models to be determined, as opposed to determining the critical value of
$\beta$ for fragmentation for a given initial disk model, which was 
the goal of the present work.

 In order to better understand the approach to gravitational
instability, a second set of eight models with the $\beta$ cooling
approximation is also currently underway. These models all start
from the same $Q_{min} = 2.7$ model as in this paper, and with same 
range of eight $\beta$ values as here, but with the minimum disk 
temperature values all relaxed to 40 K (i.e., relaxed to the outer 
disk minimum temperature for the highly unstable models with 
initial $Q_{min} = 1.3$). These models will thus study the effect
of varied cooling rates on the approach to a gravitationally unstable
phase of disk evolution.

 I thank the referee, Richard Durisen, for a number of perceptive
and constructive comments on the manuscript, and
Sandy Keiser for computer systems support. The calculations 
were performed primarily on the flash cluster at DTM and also on the 
Carnegie memex cluster at Stanford University.

\clearpage
\begin{deluxetable}{lccccccc}
\tablecaption{Initial conditions and results for the models.}
\label{tbl-1}
\tablewidth{0pt}
\tablehead{\colhead{Model} 
& \colhead{$T_o$ (K)} 
& \colhead{$Q_i$}
& \colhead{$\beta$}
& \colhead{final time (yrs)} 
& \colhead{final $N_r$}
& \colhead{final $N_\phi$}
& \colhead{maximum $N_{VP}$} }
\startdata

1.3-1   & 40 & 1.3 & 1   & 244. & 200 & 1024 &10 \\     
   
1.3-3   & 40 & 1.3 & 3   & 291. & 200 & 1024 & 5 \\     
    
1.3-10  & 40 & 1.3 & 10  & 275. & 200 & 1024 & 6 \\     
 
1.3-20  & 40 & 1.3 & 20  & 378. & 200 & 1024 & 5 \\     
   
1.3-30  & 40 & 1.3 & 30  & 302. & 200 & 1024 & 5 \\     
   
1.3-40  & 40 & 1.3 & 40  & 285. & 200 & 1024 & 4 \\     
   
1.3-50  & 40 & 1.3 & 50  & 306. & 200 & 1024 & 9 \\     
\vspace{0.1in}   
1.3-100 & 40 & 1.3 & 100 & 342. & 200 & 1024 & 5 \\

1.6-1   & 60 & 1.6 & 1   & 312. & 200 & 1024 & 3 \\     
   
1.6-3   & 60 & 1.6 & 3   & 270. & 200 & 1024 & 6 \\     
    
1.6-10  & 60 & 1.6 & 10  & 263. & 200 & 1024 & 2 \\     
    
1.6-20  & 60 & 1.6 & 20  & 319. & 200 & 1024 & 2 \\     
   
1.6-30  & 60 & 1.6 & 30  & 263. & 200 & 1024 & 1 \\     
   
1.6-40  & 60 & 1.6 & 40  & 293. & 200 & 1024 & 2 \\     
 
1.6-50  & 60 & 1.6 & 50  & 295. & 200 & 1024 & 3 \\     
\vspace{0.1in}   
1.6-100 & 60 & 1.6 & 100 & 799. & 200 & 512  & 0 \\

1.7-1   & 70 & 1.7 & 1   & 358. & 200 & 1024 & 1 \\     
  
1.7-3   & 70 & 1.7 & 3   & 327. & 200 & 1024 & 3 \\     
    
1.7-10  & 70 & 1.7 & 10  & 292. & 200 & 1024 & 4 \\     
\vspace{0.1in}      
1.7-100 & 70 & 1.7 & 100 & 393. & 200 & 1024 & 3 \\

1.9-1   & 80 & 1.9 & 1   & 394. & 200 & 1024 & 5 \\     
 
1.9-3   & 80 & 1.9 & 3   & 286. & 200 & 1024 & 3 \\     
    
1.9-10  & 80 & 1.9 & 10  & 413. & 200 & 1024 & 1 \\     
\vspace{0.1in}      
1.9-100 & 80 & 1.9 & 100 & 470. & 200 & 1024 & 1 \\

2.0-1   & 90 & 2.0 & 1   & 235. & 200 & 1024 & 3 \\     
   
2.0-3   & 90 & 2.0 & 3   & 383. & 200 & 1024 & 3 \\     
    
2.0-10  & 90 & 2.0 & 10  & 351. & 200 & 1024 & 1 \\     
\vspace{0.1in}      
2.0-100 & 90 & 2.0 & 100 & 315. & 200 & 1024 & 0 \\

2.1-1   & 100 & 2.1 & 1   & 354. & 200 & 512  & 0 \\     
   
2.1-3   & 100 & 2.1 & 3   & 486. & 200 &1024  & 0 \\     
    
2.1-10  & 100 & 2.1 & 10  & 896. & 200 &1024  & 0 \\     
\vspace{0.1in}      
2.1-100 & 100 & 2.1 & 100 & 490. & 200 & 512  & 0 \\

2.2-1   & 120 & 2.2 & 1   & 415. & 200 & 1024 & 1 \\     
   
2.2-3   & 120 & 2.2 & 3   & 697. & 200 & 512  & 0 \\     
    
2.2-10  & 120 & 2.2 & 10  & 772. & 100 & 512  & 0 \\     
\vspace{0.1in}      
2.2-100 & 120 & 2.2 & 100 & 978. & 100 & 512  & 0 \\

2.7-1   & 180 & 2.7 & 1   & 450. & 100 & 512  & 0 \\     
   
2.7-3   & 180 & 2.7 & 3   & 454. & 100 & 512  & 0 \\     
    
2.7-10  & 180 & 2.7 & 10  & 414. & 100 & 512  & 0 \\     
      
2.7-100 & 180 & 2.7 & 100 & 414. & 100 & 512  & 0 \\   

\enddata
\end{deluxetable}

\clearpage

\begin{figure}
\vspace{-2.0in}
\plotone{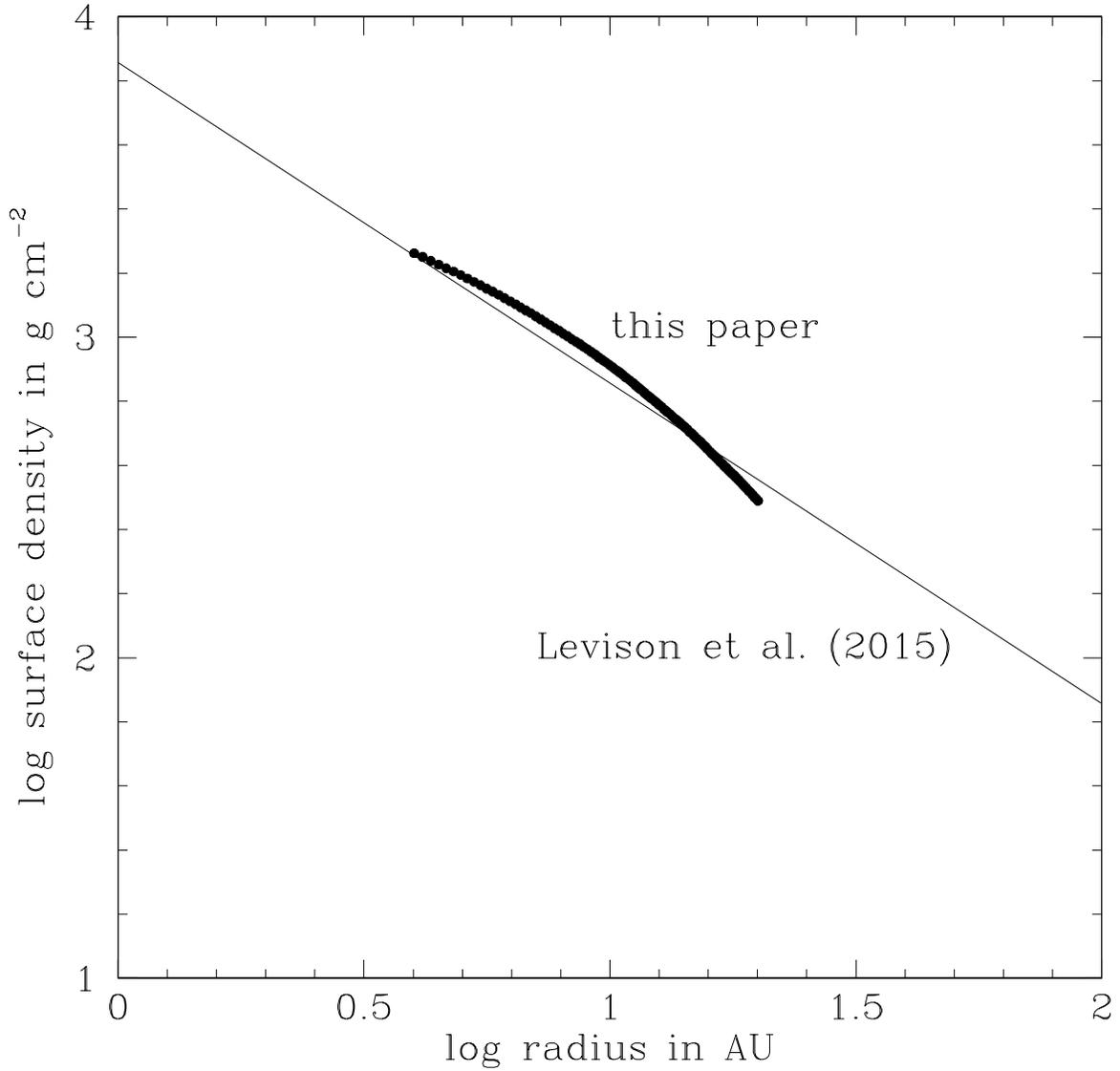}
\caption{Initial disk surface density profile for all the models, compared 
to that assumed in the core accretion models by Levison et al. (2015). 
The present model disks all have an inner radius of 4 AU and an outer 
radius of 20 AU, for a total disk mass of 0.091 $M_\odot$.}
\end{figure}

\begin{figure}
\vspace{-2.0in}
\plotone{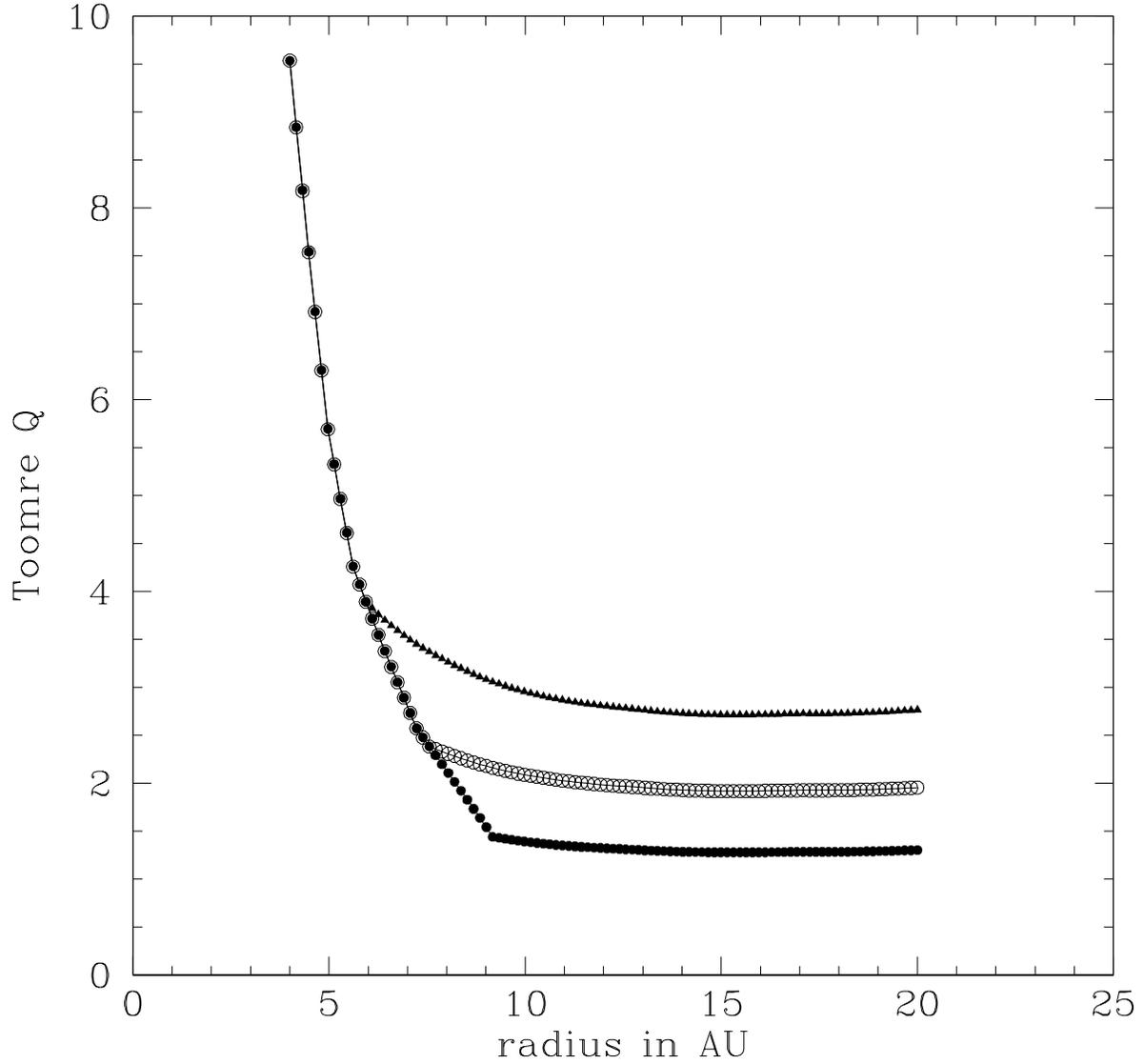}
\caption{Initial Toomre (1964) $Q$ stability parameter as a function of
disk radius for models with outer disk temperatures of 40 K, 90 K, and
180 K, leading to initial minimum $Q$ values ($Q_i$) beyond 10 AU of 1.3 
(bottom), 2.0 (middle), and 2.7 (top), respectively.}
\end{figure}

\begin{figure}
\vspace{-2.0in}
\plotone{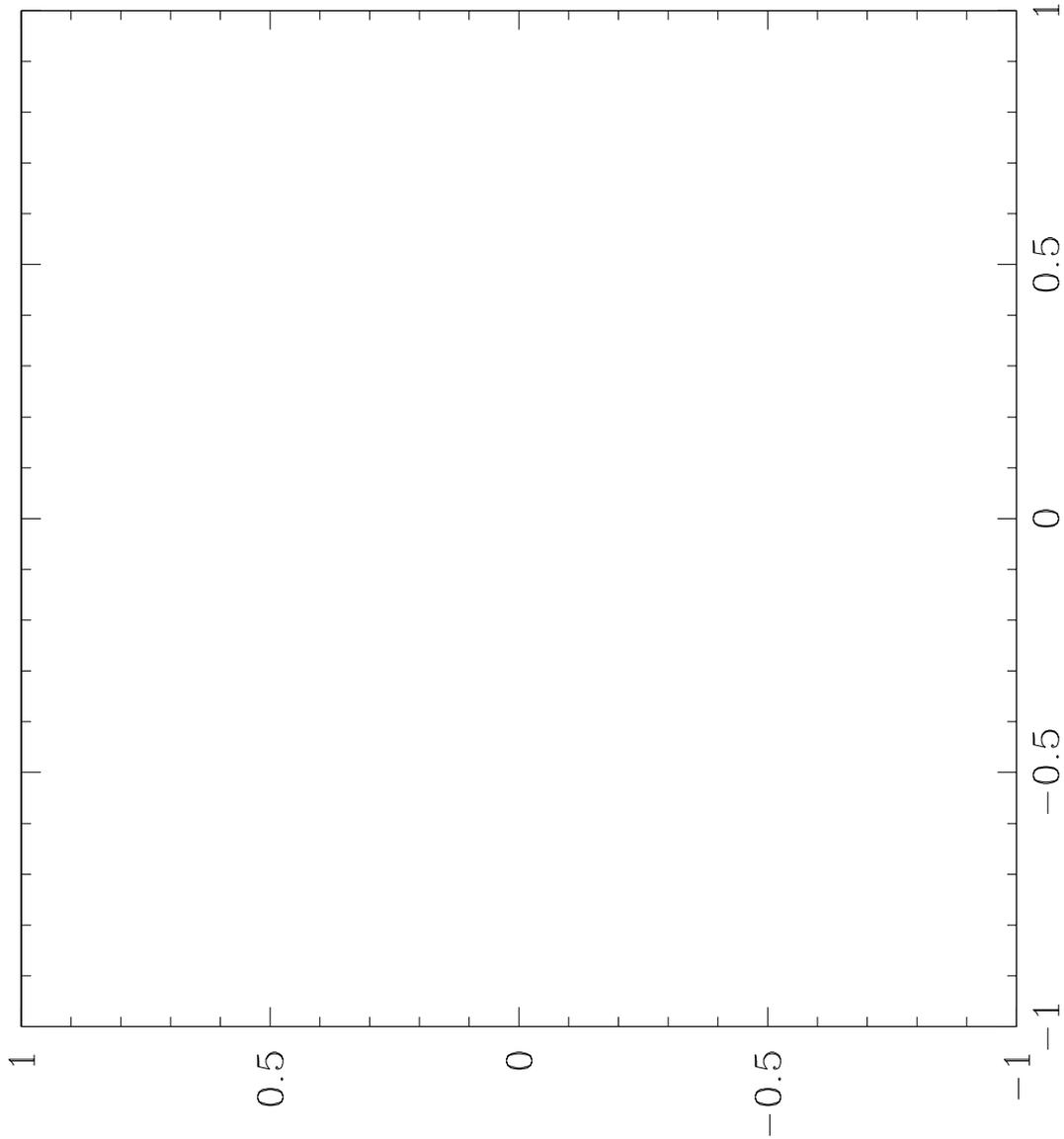}
\caption{Note: Inclusion of Figure 3 violates the file size limits for arxiv.
You may download the entire pdf file from this web page:
https://home.dtm.ciw.edu/users/boss/ftp/beta-cooling.pdf. Caption for Figure 3:
Equatorial (midplane) density (a) and temperature (b) contours 
for model 1.3-1 at the beginning of the evolution. The disk has an inner 
radius of 4 AU and an outer radius of 20 AU. Contours are labelled in 
log cgs and log K units, respectively.}
\end{figure}

\begin{figure}
\vspace{-2.0in}
\plotone{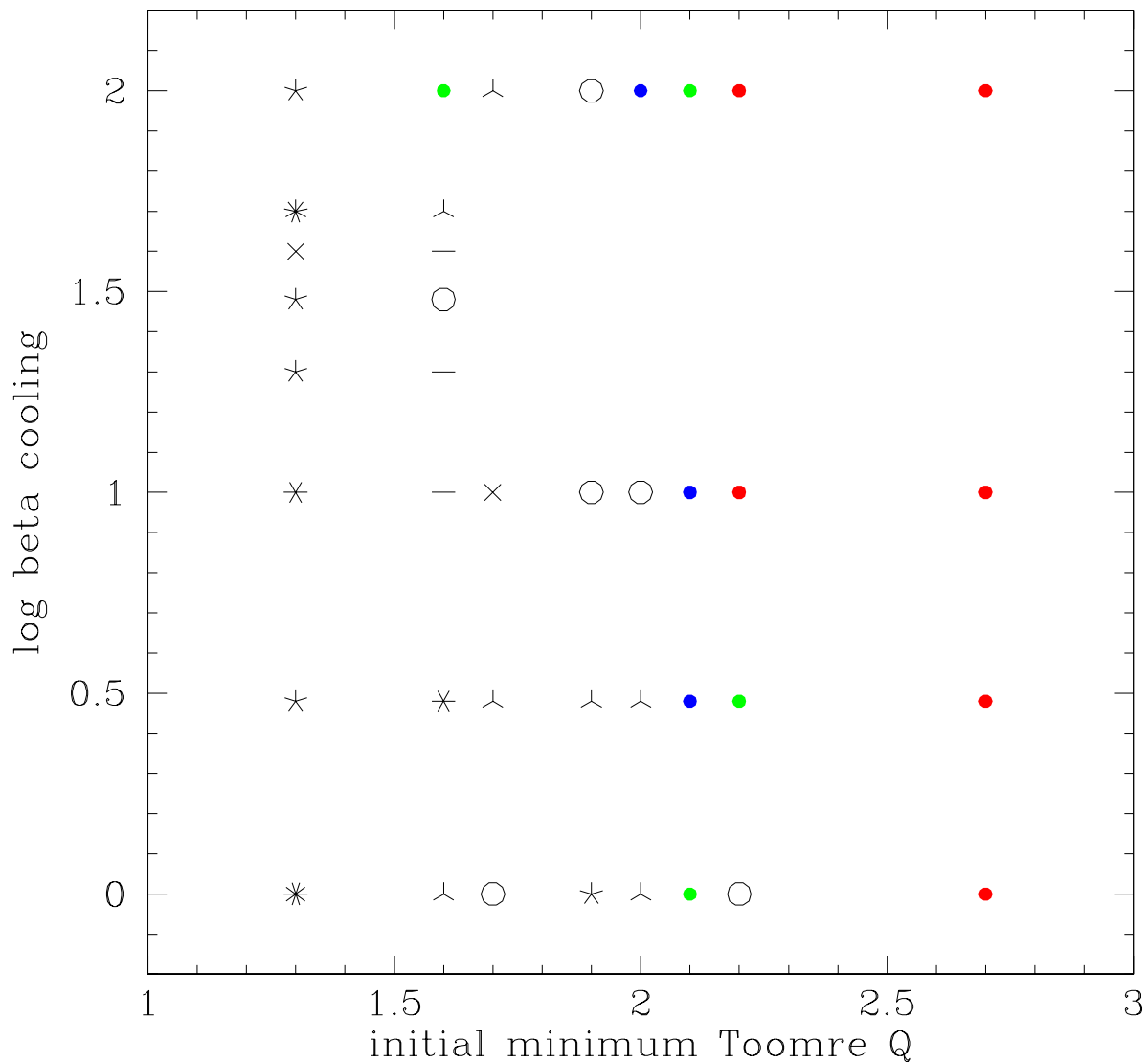} 
\caption{Results for the models, plotted as a function of the minimum of
the initial Toomre (1964) $Q$ stability parameter and the beta cooling
parameter (see Table 1). Symbols denote the degree of growth of
nonaxisymmetry or fragmentation, as follows: red = little growth, 
green = moderate growth, blue = significant growth, black = fragmentation
into virtual protoplanets (VPs), with the symbol shape denoting the maximum
number of VPs formed (e.g., circles denote single VPs, bars denote two VPs, 
six pointed-stars denote six VPs, etc.)}
\end{figure}

\begin{figure}
\vspace{-1.0in}
\plotone{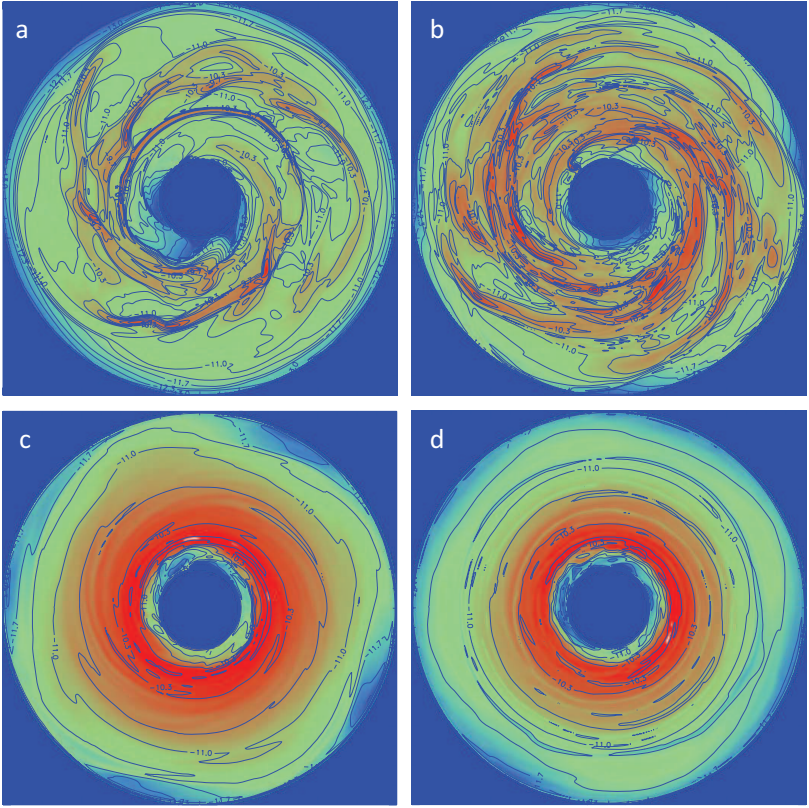}
\caption{Equatorial (midplane) density contours for (a) model 1.3-1 after 
244 yr, (b) model 1.3-100 after 342 yr, (c) model 2.7-1 after 450 yr, and
(d) model 2.7-100 after 414 yr, plotted as in Figure 3.}
\end{figure}

\begin{figure}
\vspace{-1.0in}
\plotone{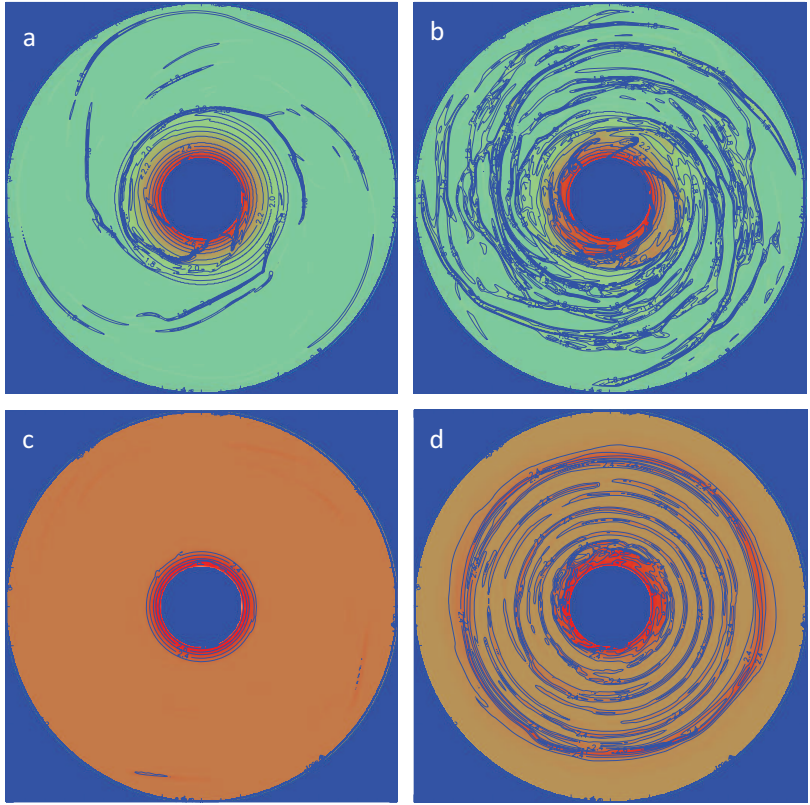}
\caption{Equatorial (midplane) temperature contours for (a) model 1.3-1 after 
244 yr, (b) model 1.3-100 after 342 yr, (c) model 2.7-1 after 450 yr, and
(d) model 2.7-100 after 414 yr, plotted as in Figure 3. Models 1.3-1 
and 1.3-100 have minimum temperatures of 40 K (light green color), 
while models 2.7-1 and 2.7-100 have minimum temperatures of 180 K
(light orange color).}
\end{figure}

\begin{figure}
\vspace{-2.0in}
\plotone{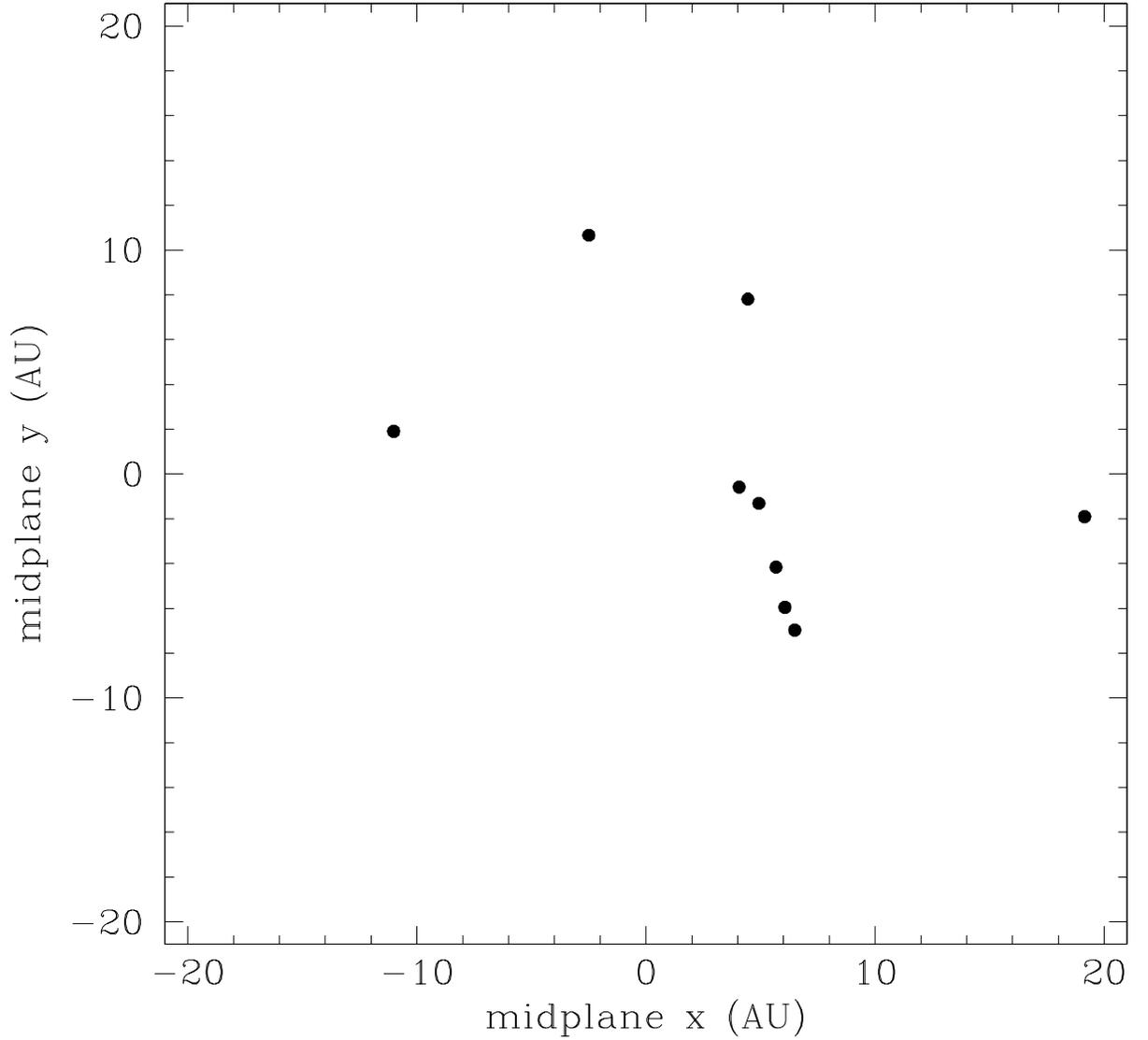}
\caption{Midplane locations of the nine VPs still active in model 1.3-1 
at the final time of 244 yr, for comparison to the midplane density
and temperature contours shown in Figures 5a and 6a, respectively. 
The VPs are not associated with any specific spiral arms due to
the ensuing chaotic evolution of both their orbits and the gas density 
maxima since the creation of the VPs.}
\end{figure}

\begin{figure}
\vspace{-2.0in}
\plotone{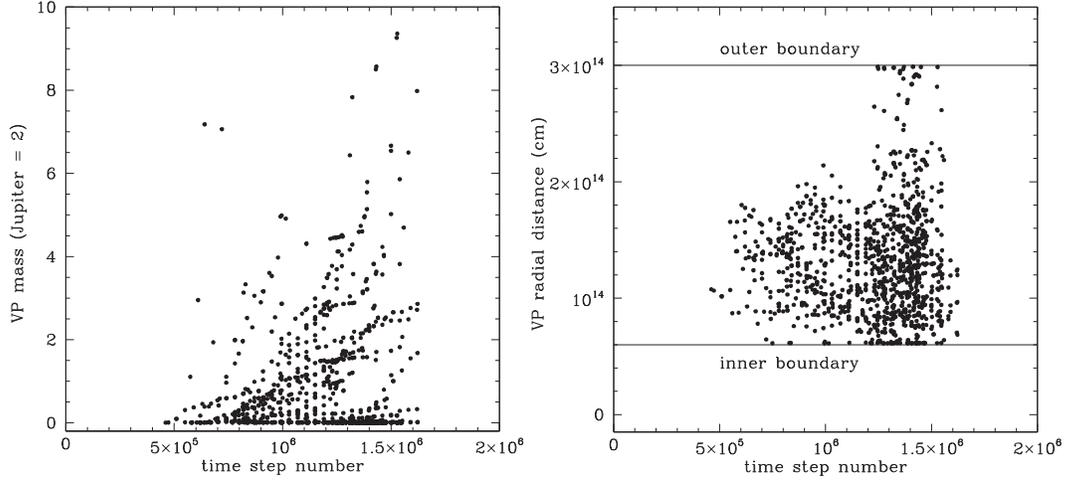}
\caption{Masses (left) and orbital radii (right) as a function of time step number
for all of the VPs formed by the eight models starting with initial minimum 
Toomre $Q = 1.3$, sampled every 40,000 time steps throughout their evolutions.
The masses increase by BHL accretion, and the orbital radii both 
increase and decrease after forming in the about 6 AU to 10 AU region. VPs that
hit the inner or outer boundaries are removed from the evolutions.}
\end{figure}

\begin{figure}
\vspace{-2.0in}
\plotone{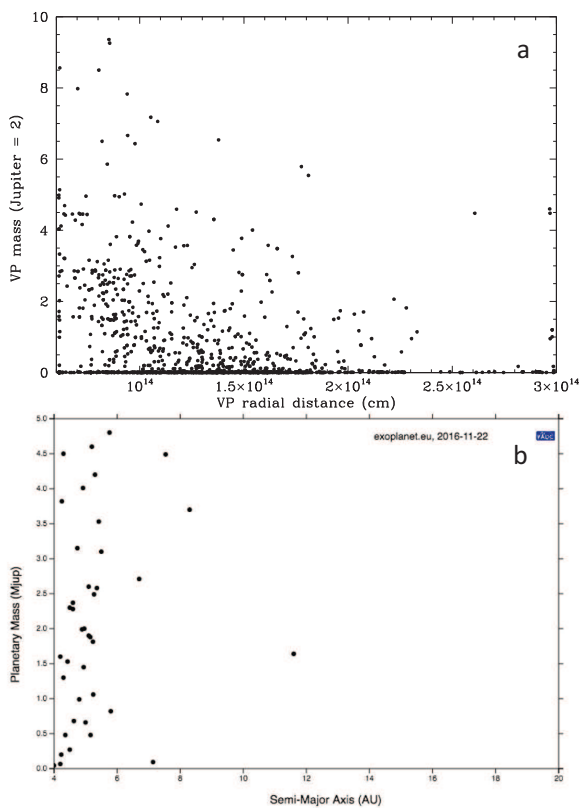}
\caption{A first look at a possible population synthesis model for gas
giants formed by disk instability (a) compared to all exoplanets (b) in the
Extrasolar Planets Encyclopedia (exoplanets.eu) as of November 22, 2016, for 
masses between 0 and 5 $M_J$ and semi-major axes between 4 AU and 20 AU.
The masses and orbital radii of all of the VPs formed by the eight models
starting with initial minimum Toomre $Q = 1.3$ are shown in (a), sampled
about 40 times throughout their evolutions. While the distributions look
similar for semi-major axes less than 6 AU, these models predict that there 
should be a significant number of gas giants with masses of about 1 $M_J$
and semi-major axes of about 6 AU to 16 AU remaining to be discovered.}
\end{figure}

\begin{figure}
\vspace{-2.0in}
\plotone{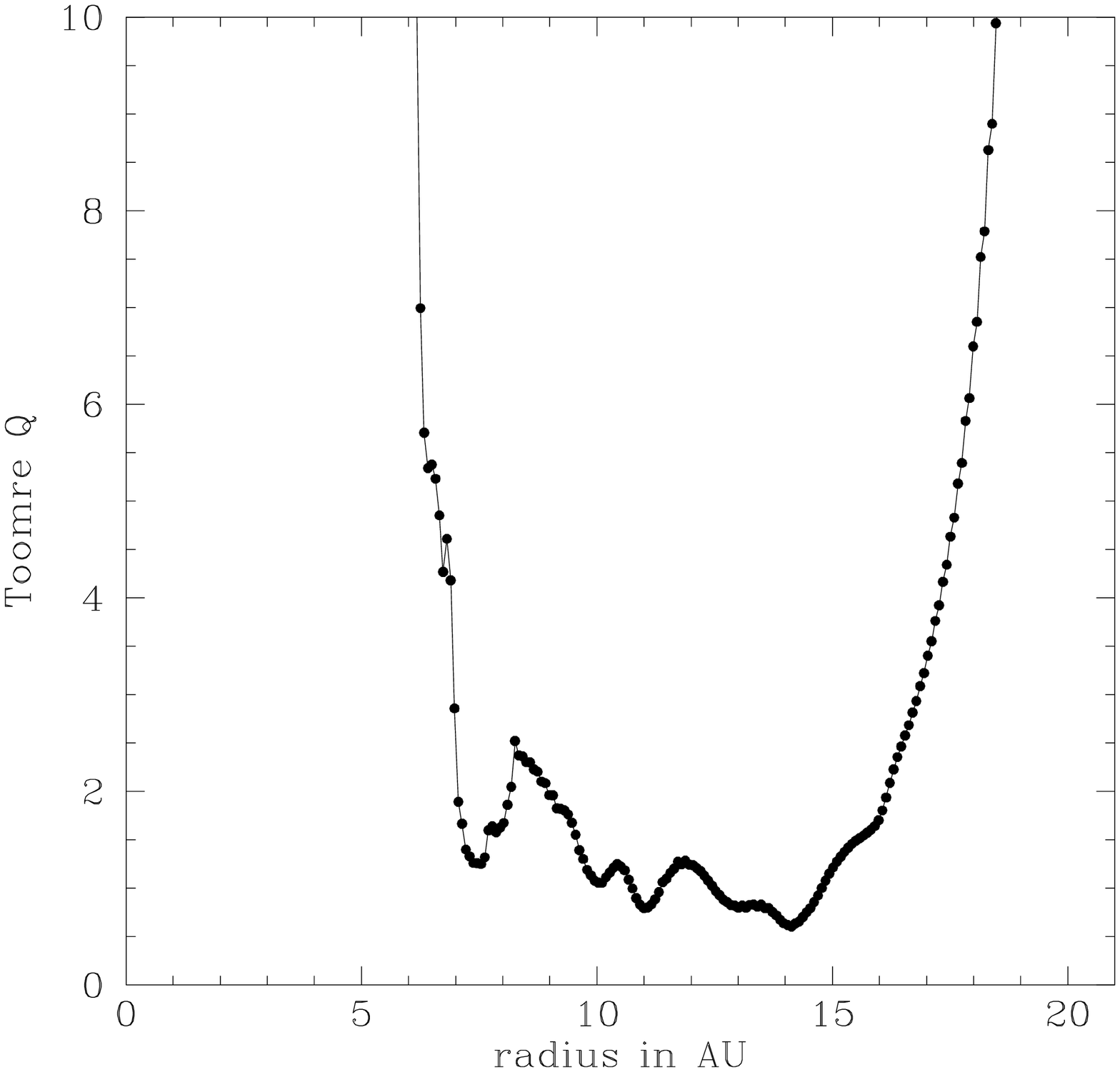}
\caption{Azimuthally averaged radial profile of the Toomre $Q$ parameter
for model 1.3-1 at a time of 104 yrs, when the disk non-axisymmetry became
so strong that a VP had to be inserted, marking the onset of fragmentation.
The minimum $Q$ value at that time is 0.61, consistent with the results
of Takahashi et al. (2016).}
\end{figure}


\begin{references}

\reference{r}
Adams, F. C., Ruden, S. P., \& Shu, F. H. 1989, ApJ,  347, 959

\reference{r}
Akiyama, E., Hasegawam Y., Hayashi, M., \& Iguchi, S. 2016, ApJ, 818, 158

\reference{r}
ALMA Partnership et al. 2015, ApJL, 808, L3

\reference{r}
Alibert, Y., Mordasini, C., \& Benz, W. 2011, A\&A, 526, A63

\reference{r}
Andrews, S. M., \& Williams, J. P. 2007, ApJ, 659, 705

\reference{r}
Andrews, S. M., Rosenfeld, K. A., Kraus, A. L., \& Wilner, D. J. 2013, 
ApJ, 771, 129

\reference{r}
Andrews, S. M., Wilner, D. J., Hughes, A. M., Qi, C., \& Dullemond, C. P.
2010, ApJ, 723, 1241

\reference{r}
Backus, I., \& Quinn, T. 2016, MNRAS, 463, 2480

\reference{r}
Baehr, H., \& Klahr, H. 2015, ApJ, 814, 155

\reference{r}
Banzatti, A., Testi, L., Isella, A., Natta, A., Neri, R., \& Wilner, D. J.
2011, A\&A, 525, A12

\reference{r}
Batalha, N. 2014, PNAS, 111, 12647

\reference{r}
Bitsch, B., Lambrechts, M., \& Johansen, A.2015, A\&A, 582, A112

\reference{r}
Boley, A. C. 2009, ApJ, 695, L53

\reference{r}
Boley, A. C., \& Durisen, R. H. 2006, ApJ, 641, 534

\reference{r}
Boley, A. C., \& Durisen, R. H. 2008, ApJ, 685, 1193

\reference{r}
Boley, A. C., \& Durisen, R. H. 2010, ApJ, 724, 618

\reference{r}
Boley, A. C., Durisen, R. H., Nordlund, A., \& Lord, J. 2007, 
ApJ, 665, 1254

\reference{r}
Boley, A. C., Hayfield, T., Mayer, L., \& Durisen, R. H. 2010, Icarus, 
207, 509

\reference{r}
Boley, A. C., Mej\'ia, A. C., Durisen, R. H., Cai, K., Pickett, M. K.,
\& D'Alessio, P. 2006, ApJ, 651, 517

\reference{r}
Borucki, W., et al. 2010, Science, 327, 977

\reference{r}
Borucki, W., et al. 2011a, ApJ, 728, 117

\reference{r}
Borucki, W., et al. 2011b, ApJ, 736, 19

\reference{r}
Boss, A. P. 1993, ApJ, 417, 351

\reference{r}
Boss, A. P. 1996, ApJ, 469, 906

\reference{r}
Boss, A. P. 1997, Science, 276, 1836

\reference{r}
Boss, A. P. 1998, ApJ, 503, 923    

\reference{r}
Boss, A. P. 2001, ApJ, 563, 367  

\reference{r}
Boss, A. P. 2002, ApJ, 567, L149  

\reference{r}
Boss, A. P. 2004, ApJ, 610, 456 

\reference{r}
Boss, A. P. 2005, ApJ, 629, 535 

\reference{r}
Boss, A. P. 2006, ApJ, 641, 1148

\reference{r}
Boss, A. P. 2007, ApJ, 661, L73 

\reference{r}
Boss, A. P. 2008, ApJ, 677, 607 

\reference{r}
Boss, A. P. 2009, ApJ, 694, 107 

\reference{r}
Boss, A. P. 2010, ApJ, 725, L145   

\reference{r}
Boss, A. P. 2011 ApJ, 731, 74    

\reference{r}
Boss, A. P. 2012, MNRAS, 419, 1930 

\reference{r}
Boss, A. P. 2013, ApJ, 764, 194   

\reference{r}
Boss, A. P. 2015, ApJ, 807, 10    

\reference{r}
Boss, A. P., \& Myhill, E. A. 1992, ApJS, 83, 311

\reference{r}
Boss, A. P., Fisher, R. T., Klein, R. I., \& McKee, C. F. 2000, ApJ, 528, 
325

\reference{r}
Bromley, B. C., \& Kenyon, S. J. 2016, ApJ, 826, 64

\reference{r}
Bryan, M. L., et al. 2016, ApJ, 821, 89

\reference{r}
Cai, K., Durisen, R. H., Boley, A. C., Pickett, M. K., \& Mejia, A. C. 
2008, ApJ, 673, 1138

\reference{r}
Cai, K., Pickett, M. K., Durisen, R. H., \& Milne, A. M. 2010, ApJ, 716, L176

\reference{r}
Cameron, A. G. W. 1978, Moon Planets, 18, 5

\reference{r}
Carrasco-Gonzalez, C., et al. 2016, ApJL, 821, L16 

\reference{r}
Cassan, A., et al. 2012, Nature, 481, 167

\reference{r}
Chambers, J. E. 2006, ApJL, 652, L133

\reference{r}
Chambers, J. E. 2016, ApJ, 825, 63

\reference{r}
Cieza, L., et al. 2007, ApJ, 667, 308

\reference{r}
Cieza, L., et al. 2015, MNRAS, 454, 1909

\reference{r}
Coleman, G. A. L., \& Nelson, R. P. 2014, MNRAS, 445, 479

\reference{r}
Coleman, G. A. L., \& Nelson, R. P. 2016, MNRAS, 460, 2779

\reference{r}
Cumming, A., et al. 2008, PASP, 120, 531

\reference{r}
Currie, T., et al. 2014, ApJL, 780, L30

\reference{r}
D’Alessio, P., Calvet, N., Hartmann, L., Franco-Hernandez, R., \&
Servin, H. 2006, ApJ, 638, 314

\reference{r} 
Dartois, E., Dutrey, A., \& Guilloteau, S. 2003, A\&A, 399, 773

\reference{r}
Dipierro, G., Price, D., Laibe, G., Hirsh, K., Cerioli, A., \&
Lodato, G. 2015, MNRAS, 453, L73

\reference{r}
Dittkrist, K.-M., Mordasini, C., Klahr, H., Alibert, Y., \& Henning, T.
2014, A\&A, 567, A121

\reference{r}
Dong, R., Hall, C., Rice, K., \& Chiang, E. 2015, ApJL, 812, L32

\reference{r}
Dunham, M. M., Vorobyov, E. I., \& Arce, H. G. 2014, MNRAS, 444, 887

\reference{r}
Durisen, R. H., Boss, A. P., Mayer, L., Nelson, A., Rice, K., \& Quinn, T. R.
2007, in Protostars and Planets V, B. Reipurth, D. Jewitt, \& K. Keil, eds. 
(Tucson: University of Arizona Press), 607
 
\reference{r}
Federrath, C., Banerjee, R., Clark, P. C., \& Klessen, R. S. 2010,
ApJ, 713, 269

\reference{r}
Forgan, D., Parker, R. J., \& Rice, K. 2015, MNRAS, 447, 836

\reference{r}
Forgan, D., \& Rice, K. 2013, MNRAS, 423, 3168

\reference{r}
Forgan, D. H., Ilee, J. D., Cyganowski, C. J., Brogan, C. L., \& Hunter, T. R. 
2016, MNRAS, 463, 957

\reference{r}
Fressin, F., et al. 2013, ApJ, 766, 81

\reference{r}
Fu, R. R., Lima, E. A., \& Weiss, B. P. 2014, EPSL, 404, 54

\reference{r}
Galvagni, M., Hayfield, T., Boley, A., Mayer, L., Roskar, R., \& Saha, P.
2012, MNRAS, 427, 1725

\reference{r}
Gammie, C. F. 2001, ApJ, 553, 174

\reference{r}
Hama, T., Kouchi, A., \& Watanabe, N. 2016, Science, 351, 65

\reference{r}
Helled, R., et al. 2014, in Protostars and Planets VI, 
H. Beuther et al., eds. (Tucson: University of Arizona Press), 643

\reference{r}
Hernandez, J., et al. 2008, ApJ, 686, 1195

\reference{r}
Hogerheijde, M. R., et al. 2011, Science, 334, 338

\reference{r}
Hubickyj, O., Bodenheimer, P., \& Lissauer, J. J. 2005, Icarus, 179, 415

\reference{r}
Ida, S., \& Lin, D. N. C. 2004, ApJ, 604, 388

\reference{r}
Ida, S., \& Lin, D. N. C. 2005, ApJ, 626, 1045

\reference{r}
Ida, S., \& Lin, D. N. C. 2008, ApJ, 685, 584

\reference{r}
Isella, A., Carpenter, J. M., \& Sargent, A. I. 2009, ApJ, 701, 260  

\reference{r}
Kaib, N., \& Chambers, J. E. 2016, MNRAS, 455, 3561

\reference{r}
Kawakita, H., et al. 2001, Science, 294, 1089

\reference{r}
Kenyon, S. J., \& Bromley, B. C. 2015, ApJ, 806, 42

\reference{r}
Kikuchi, A., Higuchi, A., \& Ida, S. 2014, ApJ, 797, 1

\reference{r}
Klassen, M., et al. 2014, 797, 4

\reference{r}
Konopacky, Q. M., et al. 2016, ApJL, 829, L4

\reference{r}
Kratter, K. M., Murray-Clay, R. A., \& Youdin, A. N. 2010, ApJ, 710, 1375

\reference{r}
Lachaume, R., Malbet, F., \& Monin, J.-L. 2003, A\&A, 400, 185

\reference{r}
Lectez, S., et al. 2015, ApJL, 805, L1

\reference{r}
Levison, H. F., Kretke, K. A., \& Duncan, M. J. 2015, Nature, 524, 322

\reference{r}
Li, Y., Kouwenhoven, M. B., Stamatellos, D., \& Goodwin, S. P. 2015,
MNRAS 805, 116

\reference{r}
Lissauer, J. J., \& Stevenson, D. J. 2007, in Protostars and Planets V, 
B. Reipurth, D. Jewitt, \& K. Keil, eds. 
(Tucson: University of Arizona Press), 591

\reference{r}
Liu, H. B., et al. 2016, Science Advances, Vol. 2, No. 2, id.e1500875

\reference{r}
Lodato, G., \& Rice, W. K. M. 2004, MNRAS, 351, 630

\reference{r}
Lyra, W., et al. 2016, ApJ, 817, 102

\reference{r}
Ma, B., \& Ge, J. 2014, MNRAS, 439, 2781

\reference{r}
Ma, S., Mao, S., Ida, S., Zhu, W., \& Lin, D. N. C. 2016, MNRAS, 461, L107

\reference{r}
Macintosh, B., et al. 2015, Science, 350, 64

\reference{r}
Madhusudhan, N., Amin, M. A., \& Kennedy, G. M. 2014, ApJL, 794, L12

\reference{r}
Mann, R. K., et al. 2015, ApJ, 802, 77

\reference{r}
Marois, C., Macintosh, B., Barman, T., Zuckerman, B., Song, I., 
Patience, J., Lafreni\'ere, D., \& Doyon, R. 2008, Science, 322, 1348

\reference{r}
Marois, C., Zuckerman, B., Konopacky, Q. M., Macintosh, B., \&
Barman, T. 2010, Nature, 468, 1080

\reference{r}
Mayer, L., Lufkin, G., Quinn, T., \& Wadsley, J. 2007, ApJ, 661, L77

\reference{r}
Mayer, L., Peters, T., Pineda, J. E., Wadsley, J., \& Rogers, P. 2016,
ApJL, 823, L36

\reference{r}
Mejia, A. C., Durisen, R. H., Pickett, M. K., \& Cai, K. 2005, ApJ, 619, 1098

\reference{r}
Meru, F. 2015, MNRAS, 454, 2529 

\reference{r}
Meru, F., \& Bate, M. R. 2010, MNRAS, 409?, 858 

\reference{r}
Meru, F., \& Bate, M. R. 2011a, MNRAS, 410, 559 

\reference{r}
Meru, F., \& Bate, M. R. 2011b, MNRAS, 411, L1 

\reference{r}
Meru, F., \& Bate, M. R. 2012, MNRAS, 427, 2022

\reference{r}
Miotello, A., Bruderer, S., \& van Dishoeck, E. F. 2014, A\&A, 572, A96

\reference{r}
Miotello, A., van Dishoeck, E. F., Kama, M., \& Bruderer, S. 2016, A\&A, 594, A85

\reference{r}
Mizuno, H. 1980, Prog. Theor. Phys., 64, 544

\reference{r}
Mordasini, C., Alibert, Y., Klahr, H., \& Henning, T. 2012, A\&A, 547, A111

\reference{r}
Mousis, O., et al. 2016, ApJL, 819, L33

\reference{r}
Nayakshin, S. 2010, MNRAS, 408, L36

\reference{r}
Nayakshin, S. 2015a, MNRAS, 446, 459

\reference{r}
Nayakshin, S. 2015b, MNRAS, 448, L25

\reference{r}
Nelson, A. F. 2006, MNRAS, 373, 1039

\reference{r}
Nero, D., \& Bjorkman, J. E. 2009, ApJ, 702, L163

\reference{r}
Paardekooper, S.-J. 2012, MNRAS, 421, 3286

\reference{r}
P\'erez, L. M., et al. 2016, Science, 353, 1519

\reference{r}
Pickett, B. K., et al. 2000, ApJ, 529,1034

\reference{r}
Pohl, A., et al. 2015, MNRAS, 453, 1768

\reference{r}
Pollack, J. B., Hubickyj, O., Bodenheimer, P., Lissauer, J. J., Podolak, M.,
\& Greenzweig, Y. 1996, Icarus, 124, 62

\reference{r}
Pringle, J. E. 1981, ARAA, 19, 137

\reference{r}
Quanz, S. P., Lafreniere, D., Meyer, M. R., Reggiani, M. M., \&
Buenzli, E. 2012, A\&A, 541, A133

\reference{r}
Quanz, S. P., Amara, A., Meyer, M. R., Girard, J. H., Kenworthy, M. A.,
\& Kasper, M. 2015, ApJ, 807, 64

\reference{r}
Rafikov, R. R. 2015, ApJ, 804, 62

\reference{r}
Rice, K., Lopez, E., Forgan, D., \& Biller, B. 2015, MNRAS, 454, 1940

\reference{r}
Rice, W. K. M., Paardekooper, S.-J., Forgan, D. H., \& Armitrage, P. J. 
2014, MNRAS, 438, 1593

\reference{r}
Rogers, P. D., \& Wadsley, J. 2012, MNRAS, 423, 1896

\reference{r} 
Ruffert, M., \& Arnett, D. 1994, ApJ, 427, 351

\reference{r}
Sallum, S., et al. 2015, Nature, 527, 342

\reference{r}
Shvartzvald, Y., et al. 2016, MNRAS, 457, 4089

\reference{r}
Stamatellos, D. 2015, ApJL, 810, L11

\reference{r}
Steiman-Cameron, T., Durisen, R. H., Boley, A. C., Michael, S., \&
McConnell, C. R. 2013, ApJ, 768

\reference{r}
Stephens, I. W., et al. 2014, Nature,514, 597

\reference{r}
Sumi, T., et al. 2011, Nature, 473, 349

\reference{r}
Takahashi, S. Z., Tsukamoto, Y., \& Inutsuka, S. 2016, MNRAS, 458, 3597

\reference{r}
Testi, L, et al. 2015, ApJL, 812,L38

\reference{r}
Toomre, A. 1964, ApJ, 139, 1217

\reference{r}
Tsukamoto, Y., Takahashi, S. Z., Machida, M. N., \& Inutsuka, S. 2015, 
MNRAS, 446, 1175

\reference{r}
Veras, D., \& Raymond, S. N. 2012, MNRAS, 421, L117

\reference{r}
Vorobyov, E. I. 2013, A\&A, 552, A129

\reference{r}
Vorobyov, E. I. 2016, A\&A, 590, A115

\reference{r}
Vorobyov, E. I., \& Basu, S. 2010a, ApJL, 714, L133

\reference{r}
Vorobyov, E. I., \& Basu, S. 2010b, ApJ, 719, 1896

\reference{r}
Weidenschilling, S. J. 1977, ApSS, 51, 153

\reference{r}
Wetherill, G. W. 1990, AREPS, 18, 205

\reference{r}
Wetherill, G. W. 1996, Icarus, 119, 219

\reference{r}
Williams, J. P., et al. 2013, MNRAS, 435, 1671

\reference{r}
Wittenmyer, R. A., et al. 2016, ApJ, 819,28

\reference{r}
Young, M. D., \& Clarke, C. J. 2015, MNRAS, 451, 3987

\reference{r}
Young, M. D., \& Clarke, C. J. 2016, MNRAS, 455, 1438

\reference{r}
Zhu, Q., Hernquist, L., \& Li, Y. 2015, ApJ, 800, 6

\reference{r}
Zhu, Z., et al. 2007, ApJ, 669, 483

\reference{r}
Zhu, Z., Hartmann, L., Gammie, C., \& McKinney, J. C. 2009, ApJ, 701, 620

\reference{r}
Zhu, Z., Hartmann, L., \& Gammie, C. 2010, ApJ, 713, 1143


\end{references}
\end{document}